\documentclass{iopart}

	\usepackage{iopams}
	\usepackage{setstack}
		\usepackage[utf8]{inputenc}	
	\usepackage{graphicx}
	\usepackage{subfigure}

\usepackage[dvips]{color} 
	
\begin{document}

\title{Relativistic Gravothermal Instabilities}
\author{Zacharias Roupas}
\address{Institute of Nuclear and Particle Physics, N.C.S.R. Demokritos, GR-15310 Athens, Greece} 
\ead{roupas@inp.demokritos.gr}

\begin{abstract}
The thermodynamic instabilities of the self-gravitating, classical ideal gas are studied in the case of static, spherically symmetric configurations in General Relativity taking into account the Tolman-Ehrenfest effect. One type of instabilities is found at low energies, where thermal energy becomes too weak to halt gravity and another at high energies, where gravitational attraction of thermal pressure overcomes its stabilizing effect. These turning points of stability are found to depend on the total rest mass $\mathcal{M}$ over the radius $R$. The low energy instability is the relativistic generalization of Antonov instability, which is recovered in the limit $G\mathcal{M} \ll R c^2$ and low temperatures, while in the same limit and high temperatures, the high energy instability recovers the instability of the radiation equation of state. In the temperature versus energy diagram of series of equilibria, the two types of gravothermal instabilities make themselves evident as a double spiral! The two energy limits correspond also to radius limits. So that, stable static configurations exist only in between two marginal radii for any fixed energy with negative thermal plus gravitational energy. Ultimate limits of rest mass, as well as total mass-energy, are reported. Applications to neutron cores are discussed.
\end{abstract}

\maketitle

\section{Introduction}

\indent	An intriguing result of Newtonian Gravity is that a bounded sphere of ideal gas in thermal equilibrium, namely an `isothermal sphere', becomes unstable at radii \textit{greater than a maximum critical radius} for a fixed negative, gravitational plus thermal (gravothermal) energy. For a fixed radius the isothermal sphere becomes unstable for energies \textit{less than some critical value}. This instability is called Antonov instability or gravothermal catastrophe \cite{Antonov:1962,Bell:1968,Padman:1989,Axenides:2012bf}. It is completely counter-intuitive to the General Relativity paradigm, where one expects to acquire instabilities at \textit{small radii} and \textit{high masses}. The questions raised are: `how do these two opposite behaviours -newtonian and relativistic- fit into a common frame?' and `what is the relativistic generalization of Antonov instability?'.

An attempt to answer the second question was made by Chavanis \cite{Chavanis:2001ah}. However he was able to incorporate the relativistic gas only into the Newtonian Gravity and not into General Relativity. He also studied the case of the radiation equation of state in General Relativity, but this is only just a limiting case of the ideal gas equation of state, and its role will be revealed shortly. 

In the present work the full thermodynamic stability analysis of the classical ideal gas in General Relativity will be developed in the spherically symmetric static case. 
For an ideal gas, there is only thermal energy that can halt gravitational collapse. However, since thermal energy gravitates as well \cite{Tolman:1930}, it can also cause gravitational collapse at high energies. Hence, we anticipate to find two gravitational instabilities for a bounded sphere that contains an ideal gas. The system becomes unstable both at low energies, in which case I call the instabilities `\textit{the low energy gravothermal instabilities}', \textit{and} at high energies in which case I call the instabilities `\textit{the high energy gravothermal instabilities}'. Most times I use plural for each type, because at each one correspond instabilities in the various thermodynamic ensembles (ensembles in Gravity are not equivalent \cite{Padman:1990}).
\textit{The low energy gravothermal instability in the microcanonical ensemble is the relativistic generalization of Antonov instability}. 

The turning point of stability for the relativistic instabilities, will be found to depend on the total rest mass $\mathcal{M}$, over the radius $R$ of the gas sphere, i.e. on the control parameter:
\begin{equation}
	\xi = \frac{2G\mathcal{M}}{Rc^2}.
\end{equation}
It is reported an ultimate upper limit of $\xi$ for which equilibria exist. 
In the lower limit $\xi \rightarrow 0$, the Antonov instability is recovered for low temperatures $kT \ll mc^2$ and the instability of radiation is recovered for high temperatures $mc^2 \ll kT$, where $m$ is the mass of one particle. 
For every allowed value of $\xi$ both gravothermal instability types are present and make themselves evident in the temperature versus energy series of equilibria diagram as a double spiral! This double spiral of equilibria is shrinking as the relativity control parameter $\xi$ is increasing and disappears completely for $\xi > \xi_{max}$.
Thus, when General Relativity is taken into account, isothermal spheres get more unstable. 

Let me clarify that the term `isothermal spheres' in the present context of General Relativity refers to static, spherical, bounded configurations of relativistic ideal gas in thermal equilibrium. The three conditions for static thermal equilibrium in General Relativity for any equation of state are outlined in \cite{Roupas:2014nea,Roupas:2013nt}. The first is to satisfy the well-known Tolman-Oppenheimer-Volkoff (TOV) equation \cite{Tolman:1939,Oppenheimer:1939}, the second is Tolman's relation \cite{Tolman:1930,Tolman-Ehrenfest:1930}:
\begin{equation}\label{eq:Tolman}
		T(r)\sqrt{g_{tt}} = \tilde{T} \equiv const.
\end{equation}
where $\tilde{T}$ is called Tolman temperature, $T(r)$ is the local proper temperature, and $g_{tt}$ is the time-time component of the metric, and the third is the relation
\begin{equation}\label{eq:mu_Tolman}
		\mu(r)\sqrt{g_{tt}} = \tilde{\mu} \equiv const.
\end{equation}
where $\mu$ is the relativistic proper chemical potential. These equations were also derived by Klein \cite{Klein:1949} in a different way by use of the first law of thermodynamics (see also \cite{Merafina:1989}). I stress that these two relations are of thermodynamic origin, as they follow from the extremization of entropy and cannot be derived from Einstein's equations alone. They express the fact that `heat has weight' so that it rearranges itself (leading to a distribution $T(r)$) in order to counterbalance its own gravitational attraction.

An application regarding neutron stars is discussed in the last section. Although cores of neutron stars are believed to be cold and therefore completely degenerate, pure theoretical query to quantify the ability of thermal energy to halt gravitational collapse, along with the fact that protoneutron stars are in fact hot \cite{Burrows:1986me,Prakash:1997}, motivated me to calculate the upper mass limit of non-degenerate neutron cores for which the classical ideal gas equation of state applies. Practically, this calculation accounts for considering only thermal pressure and no degeneracy pressure (or nuclear interactions), in discrepancy to the standard practice regarding neutron cores. However, as we will see, the result matches so nicely observations, that one is seriously wondered if it is only a coincidence. Several issues are raised.

Let me give a brief outline of this work. In section \ref{sec:EOS} will be discussed the equation of state of the relativistic ideal gas and the conditions for thermal equilibrium. In section \ref{sec:GI} will be presented the results of numerical analysis. In section \ref{sec:NS} will be discussed implications regarding neutron stars, while section \ref{sec:Con} is devoted to conclusions.

\section{The Equation of State and Thermal Equilibrium}\label{sec:EOS}

\subsection{Classical derivation of the equation of state}\label{sec:EOS_cl}

Let briefly recall the well-known equation of state of the relativistic ideal gas in special relativity, following mainly Chandrashekhar \cite{Chandra:1938}. The generalization to General Relativity in the static case is straightforward.

The proper phase space density $f(\vec{r},\vec{p})$ of an ideal gas should satisfy the Boltzmann distribution:
\begin{equation}\label{eq:f_initial}
	f(\vec{r},\vec{p}) = A(\vec{r})e^{-\beta \epsilon}
\end{equation}
where $\beta = 1/kT$ is the inverse temperature and $\epsilon$, $\vec{p}$ the energy and momentum of one particle, respectively. This distribution can be easily verified that maximizes the Boltzmann entropy \cite{Chavanis:2001hd}.
The particle density $n(\vec{r})$ should then be given by:
\begin{equation}\label{eq:nu_basic}
	n(\vec{r}) = \int f(\vec{r},\vec{p}) d^3\vec{p}
\end{equation}
The relativistic energy of one particle is:
\begin{equation}
	\epsilon = mc^2 \left( 1 + \frac{p^2}{m^2c^2} \right)^{\frac{1}{2}} 
\end{equation}
where $m$ is the particle mass. Applying the Juettner transformation:
\begin{equation}\label{eq:Juettner}
	\frac{p}{mc} = \sinh\theta
\end{equation}
the particle energy becomes:
\begin{equation}\label{eq:e_theta}
	\epsilon = mc^2 \cosh\theta .
\end{equation}
Let introduce the modified Bessel functions:
\begin{equation}\label{eq:bessel}
	K_n(b) = \int_0^{\infty} e^{-b \cosh\theta}\cosh(n\theta)d\theta,	
\end{equation}
where $b$ is the dimensionless equivalent to inverse temperature:
\begin{equation}\label{eq:b}
	b = \frac{mc^2}{kT}
\end{equation}
Assuming spherical symmetry and applying equations (\ref{eq:f_initial}), (\ref{eq:e_theta}) and (\ref{eq:bessel}), the integral (\ref{eq:nu_basic}) may be calculated, giving $A$. Substituting $A$ into (\ref{eq:f_initial}) we get:
\begin{equation}
	f(r,p) = \frac{1}{4\pi m^3 c^3}\frac{b}{K_2(b)}n(r) e^{-b \cosh\theta }
\end{equation}
To calculate the pressure $P$, we use the formula \cite{Chandra:1938}:
\begin{equation}
	P(r) = \frac{1}{3}\int f(r,p) p\frac{\partial \epsilon}{\partial p} d^3\vec{p}
\end{equation}
After several integrations by parts and using standard identities, we finally get:
\begin{equation}\label{eq:eos_n}
	P = mc^2\frac{n}{b}
\end{equation}
It is evident, that this is the equation of state of an ideal gas. \\

The thermal energy density:
\begin{eqnarray}
	\varepsilon_T &\equiv \int_0^\infty f \epsilon d^3\vec{p},
\end{eqnarray}
 may also be calculated using standard identities and the recursion formula:
\begin{equation}\label{eq:recur}
	K_{n+1}(b) - K_{n-1}(b) = \frac{2n}{b}K_n(b)
\end{equation}
We finally get:
\begin{equation}\label{eq:e_T}
	\varepsilon_T = mc^2 n \mathcal{F}(b)
\end{equation}
where:
\begin{equation}\label{eq:F}
		 \mathcal{F}(b) = \frac{K_1(b)}{K_2(b)} + \frac{3}{b} - 1
\end{equation}

Let us now consider General Relativity. First of all, recall that, as proved by Tolman, the temperature is not homogeneous at the equilibrium, so that $\beta = \beta(\vec{r})$ \cite{Tolman:1930} (see also \cite{Roupas:2014nea,Roupas:2013nt} for a more general framework in the static case and \cite{Green:2013ica} for the stationary case). Let also recall the expressions for various forms of energy. Weinberg \cite{Weinberg} has shown that the `material' energy, that is only the energy of matter leaving out gravitational field's, is given by the integral of mass density $\rho(r)$ over the proper volume:
\begin{equation}
	M_{mat} = \int_0^R \rho(r) \,{g_{rr}}^{\frac{1}{2}} 4\pi r^2 dr
\end{equation}
while the thermal energy $E_T$ and the rest mass $\mathcal{M}$ are given by:
\begin{eqnarray}
	&E_T = \int_0^R \varepsilon_T(r) \,{g_{rr}}^{\frac{1}{2}} 4\pi r^2 dr \\
	&\mathcal{M} = \int_0^R m n(r) \,{g_{rr}}^{\frac{1}{2}} 4\pi r^2 dr 
\end{eqnarray}
Note that the total energy, including gravitational field energy, is \cite{Weinberg}:
\begin{equation}\label{eq:mass_total}
	M = \int_0^R \rho(r)\, 4\pi r^2 dr
\end{equation}
Now, by definition:
\begin{equation}
	M_{mat}c^2 = \mathcal{M}c^2 + E_T 
\end{equation}
which gives as expected:
\begin{equation}\label{eq:rho_relation}
	\rho(r) = m n(r) + \varepsilon_T/c^2 
\end{equation}
and thus
\begin{equation}
	\rho(r) = (1+\mathcal{F}(b))m n(r)
\end{equation}
Comparing the last equation with equation (\ref{eq:eos_n}) we finally derive the equation of state of the relativistic ideal gas in General Relativity:
\begin{equation}\label{eq:eos}
	P(r) = \frac{1}{b(r)(1+\mathcal{F}(b(r)))}\rho(r) c^2,
\end{equation}
where $b$ and $\mathcal{F}$ are given by equations (\ref{eq:b}) and (\ref{eq:F}), respectively.

Using the asymptotic behaviour of the modified Bessel functions (see for example \cite{Chandra:1938}), we get the values of $\mathcal{F}$ at the Newtonian $b\rightarrow \infty$ and ultra-realtivistic $b\rightarrow 0$ limits:
\begin{eqnarray}
	&\lim_{b\rightarrow \infty} \mathcal{F}(b) = \frac{3}{2b} \; &, \; \mbox{Newtonian limit} \\
	&\lim_{b\rightarrow 0} \mathcal{F}(b) = \frac{3}{b} \; &, \; \mbox{Ultra-relativistic limit} 
\end{eqnarray}
which give when applied to (\ref{eq:e_T}) and (\ref{eq:eos}):
\begin{eqnarray}
\label{eq:eos_limitN}
	&P \rightarrow \frac{\rho}{m}kT \quad \mbox{and} \quad \varepsilon_T \rightarrow \frac{3}{2}nkT \quad  &,\; \mbox{Newtonian limit} \; b\rightarrow \infty \\
\label{eq:eos_limitR}
	&P \rightarrow \frac{\rho c^2}{3}\quad \mbox{and} \quad \varepsilon_T \rightarrow 3 nkT \quad &,\;\mbox{Ultra-relativistic limit} \; b\rightarrow 0.
\end{eqnarray}

I stress that Israel \cite{Israel:1963} has proved beyond any doubt, that an ideal gas, i.e. neglecting interactions, satisfies the equation of state (\ref{eq:eos}) in General Relativity. He derived the result by maximizing the entropy in the framework of kinetic theory in General Relativity in the most general possible set-up. 

\subsection{Classical limit of the quantum gas and chemical potential}

In order to identify the conditions under which the classical limit is valid, it should be realized that the equation of state (\ref{eq:eos}) is the non-degenerate limit of the relativistic quantum ideal gas. For a quantum ideal gas, the energy distribution of one particle is given by the Fermi-Dirac or Bose-Einstein distributions for fermions or bosons respectively:
\begin{equation}\label{eq:quantum_dis}
	g(\epsilon) = \frac{1}{e^{\beta(\epsilon - \mu)}\pm 1} \;,\;
	\left\lbrace
	\begin{array}{l}
		(+)\;\mbox{for fermions} \\[2ex]
		(-)\;\mbox{for bosons}
	\end{array}
	\right.
\end{equation}
where $\epsilon$ is the energy of one particle, including rest mass in the relativistic case, and $\mu$ the chemical potential. Using the Juettner transforamtion (\ref{eq:Juettner}) and the relativistic definition of energy \begin{equation}\label{eq:e_p}
	\epsilon = \sqrt{m^2c^4 + p^2 c^2},
\end{equation}
the distribution (\ref{eq:quantum_dis}) may be written in terms of $\theta$:
\begin{equation}
	g(\theta) = \frac{1}{e^{b (\cosh\theta - \hat{\mu})}\pm 1} 
\end{equation}
where $b = mc^2/kT$ and 
\begin{equation}
	\hat{\mu} = \frac{\mu}{mc^2}. 
\end{equation}
Let us focus on the case of baryons and electrons (fermions) in which case the pressure $P$, number density $n$ and mass density $\rho$ may be written as \cite{Chandra:1938}:
\begin{eqnarray}
\label{eq:P_Q} 	
&P = \frac{8\pi m^4 c^5}{3h^3}\int_0^\infty \frac{\sinh^4\theta d\theta}{e^{b (\cosh\theta - \hat{\mu})}+1}
\\
\label{eq:n_Q}
&n = \frac{8\pi m^3 c^3}{h^3}\int_0^\infty \frac{\sinh^2\theta \cosh\theta  d\theta}{e^{b (\cosh\theta - \hat{\mu})}+1}
\\
\label{eq:rho_Q}
&\rho = \frac{8\pi m^4 c^3}{h^3}\int_0^\infty \frac{\sinh^2\theta \cosh^2\theta  d\theta}{e^{b (\cosh\theta - \hat{\mu})}+1}.
\end{eqnarray}
Let
\begin{equation}\label{eq:alpha}
	\alpha \equiv \mu\beta = \hat{\mu}b.
\end{equation}
The chemical potential is the amount of free energy 
\begin{equation}\label{eq:free}
F = E-TS
\end{equation} 
needed to give (or take) to (from) a system in order to add one particle under conditions of constant temperature. A negative chemical potential means that the system is receptive to adding particles (no external work needed), while a positive chemical potential means external work is needed. For classical systems, the chemical potential is negative because although the energy might be slightly increased contributing a small amount to plus sign of equation (\ref{eq:free}), the entropy is hugely increased, because the number of available configurations increases greatly, contributing a big amount to the minus sign of equation (\ref{eq:free}). However, for quantum systems of fermions that are completely degenerate, adding one particle increases very slightly the entropy because the energy of the particle is certain. Due to the Pauli principle, it will occupy the highest available energy level, identified with Fermi energy $\epsilon_F$ in this case.

We have the following limits:
\begin{eqnarray}
\label{eq:limit_Q}		&a \rightarrow +\infty : \mbox{ Ultra-degenerate limit} \\
\label{eq:limit_C}		&a \rightarrow -\infty : \mbox{ Classical limit} 
\end{eqnarray}

In the first case (\ref{eq:limit_Q}), the chemical potential is positive and large compared to the temperature. Let us call it $\mu = \epsilon_F$ in this case. The distribution function (\ref{eq:quantum_dis}), with the positive sign corresponding to fermions, becomes:
\begin{equation}
	g(\epsilon) \overset{\alpha \rightarrow \infty}{\longrightarrow}\left\lbrace
	\begin{array}{l}
		1, \;\epsilon \leq \epsilon_F \\[2ex]
		0, \;\epsilon > \epsilon_F
	\end{array}
	\right.
\end{equation}
Thus, the integrals (\ref{eq:P_Q}-\ref{eq:rho_Q}) have an upper limit $p_F$. The chemical potential $\mu=\epsilon_F$ is just the Fermi energy. 

In the second case, the chemical potential is large and negative $-\mu \gg kT$. The unity in the distribution function (\ref{eq:quantum_dis}) is much smaller than the exponential and can be omitted, leading to the Boltzmann distribution
\begin{equation}
	g(\epsilon) \overset{\alpha \rightarrow -\infty}{\longrightarrow} e^{-\beta (\epsilon - \mu)}.
\end{equation}
The equations (\ref{eq:P_Q}),(\ref{eq:n_Q}) and (\ref{eq:rho_Q}), using equations (\ref{eq:bessel}), (\ref{eq:F}), and (\ref{eq:recur}) may be written in terms of the modified Bessel functions:
\begin{eqnarray}
\label{eq:P_clasB} 	
&P = \frac{8\pi m^4 c^5}{3h^3}e^{\alpha}\frac{K_2}{b^2}
\\
\label{eq:n_clasB}
&n = \frac{8\pi m^3 c^3}{h^3}e^{\alpha}\frac{K_2}{b}
\\
\label{eq:rho_clasB}
&\rho = \frac{8\pi m^4 c^3}{h^3}e^{\alpha}\frac{K_2}{b}(1+\mathcal{F})
\end{eqnarray}
Apparently, equations (\ref{eq:P_clasB}) and (\ref{eq:rho_clasB}) lead to the equation of state (\ref{eq:eos}). 

\subsection{Thermal equilibrium}\label{sec:TE}

Let discuss the conditions for thermal equilibrium in General Relativity restricting ourselves to the static spherically symmetric case for which the metric may be written as:
\begin{equation}
	ds^2 = g_{tt}dt^2 - g_{rr} dr^2 - r^2 d\Omega
\end{equation}
Since we consider an ideal gas, the energy-momentum tensor is the one of a perfect fluid, which at the equilibrium is simply:
\begin{equation}
	T^\mu_\nu = \mbox{diag}(\rho c^2,-p,-p,-p)
\end{equation}
The Einstein's equations, after some elaboration, reduce only to two equations, namely the Tolman-Oppenheimer-Volkoff (TOV) equation \cite{Tolman:1939,Oppenheimer:1939}:
\begin{equation}\label{eq:TOV}  
	\frac{d P}{dr} = - \left({\frac{P}{c^2}} + \rho\right) {\left(\frac{G\hat{M}}{r^2} + 4\pi G \frac{P}{c^2} r\right) \left(1 - \frac{2G\hat{M}}{rc^2} \right)^{-1} } 
\end{equation}
and the definition of mass
\begin{equation}\label{eq:mass_d}  
 	\frac{d \hat{M}}{dr} = 4\pi \rho r^2,	
\end{equation}
where $\hat{M}(r)$ is the total mass (rest mass+thermal energy+gravitational field's energy) until point $r$ of the sphere. We denote $R$ the edge of the gas sphere and the total mass $M=\hat{M}(R)$, as in equation (\ref{eq:mass_total}).

In Refs. \cite{Roupas:2014nea,Roupas:2013nt}, I have derived the conditions for thermal equilibrium in General Relativity for static spherically symmetric systems by extremizing the total entropy
(for the use of entropic principle see also \cite{Cocke:1965,SWZ:1981,Gao:2011hh,Gao:2012add,Green:2013ica,Fang:2013oka}) for fixed total energy and number of particles
\begin{equation}\label{eq:dS}
	\delta S - \tilde{\beta}\delta M c^2 + \alpha \delta N = 0.
\end{equation}
This condition leads to the TOV equation, assuming only the Hamiltonian constraint, which practically accounts for the mass equation (\ref{eq:mass_d}). The Lagrange multiplier $\tilde{\beta}$ is proven to be the Tolman inverse temperature, as in equation (\ref{eq:Tolman}), and Lagrange multiplier $\alpha$ is found to be exactly the quantity of equation (\ref{eq:alpha}), namely:
\begin{equation}\label{eq:alpha_cond}
	\alpha \equiv \frac{\mu(r)}{kT(r)} = const.
\end{equation}
where $T(r)$ and $\mu(r)$ are the proper temperature and chemical potential respectively. Hence, at thermal equilibrium the condition (\ref{eq:alpha_cond}) should hold. This accounts for equation (\ref{eq:mu_Tolman}).

The Lagrange multiplier $\tilde{\beta}$ identified as the Tolman temperature gives the other condition of thermal equilibrium by the definition (\ref{eq:Tolman}). In addition, equation (\ref{eq:dS}) with $\tilde{\beta}$ identified as the Tolman temperature shows that Tolman temperature is the conjugate variable to total energy. Hence, it is the quantity kept constant in the canonical ensemble. The differentiated form of equation (\ref{eq:Tolman}) may easily be calculated \cite{Roupas:2014nea}, using Einstein's equations, to be:
\begin{equation}\label{eq:tolman}  
	\frac{db}{dr} =  	-\frac{b}{P+\rho c^2} \frac{dP}{dr}, 	
\end{equation}
where $b=mc^2/kT(r)$ is the dimensionless inverse temperature. Thus, at thermal equilibrium three conditions should hold: the TOV equation, the relation (\ref{eq:alpha_cond}) and Tolman relation (\ref{eq:tolman}).

Tolman relation expresses the fact that heat has mass and therefore gravitates. At equilibrium, a temperature gradient should form to balance the gravitational attraction of heat.

The relation (\ref{eq:alpha_cond}) will enable us to calculate the chemical potential in the following section. The chemical potential, as defined here, includes the particle rest mass. It is common to redefine the chemical potential for (special) relativistic systems by subtracting the particle rest mass energy from both particle energy (\ref{eq:e_p}) and the chemical potential, and thus leaving the quantum distribution function (\ref{eq:quantum_dis}) intact. However, in General Relativity as we will see here, this cannot be done. That is because the correct chemical potential has to satisfy equation (\ref{eq:alpha_cond}). Only one definition --either with or without the rest mass-- can satisfy equation (\ref{eq:alpha_cond}) and not both. Let us prove that equations (\ref{eq:P_clasB}-\ref{eq:rho_clasB}), which express the equation of state we will use, do satisfy conditions of thermal equilibrium (\ref{eq:alpha_cond}), (\ref{eq:tolman}), with the chemical potential including the particle rest mass. The proof for the quantum expressions (\ref{eq:P_Q}-\ref{eq:rho_Q}) is given in another work \cite{Roupas:2014hqa}.

We consider the definition (\ref{eq:alpha}) for $\alpha$, namely $\alpha = \beta \mu$ and not $\alpha = \beta (\mu - mc^2)$, which would account for subtracting the rest mass. 
Let $D = \frac{8\pi m^4 c^5}{3h^3}e^{\alpha}$. Then, the pressure (\ref{eq:P_clasB}) and the density (\ref{eq:rho_clasB}) are written as:
\begin{eqnarray}
	&P = D\frac{K_2(b)}{b^2} \\
	&\rho c^2 = D\frac{K_2(b)}{b}(1+\mathcal{F}(b))
\end{eqnarray}
We assume 
\begin{equation}
	D = const.
\end{equation} 
and hence that the relation (\ref{eq:alpha_cond}) holds. 

Let calculate $dP/dr$. The derivative of $K_2$ may be calculated using the recursive formula:
\begin{equation}
	\frac{dK_n}{db} = -K_{n-1} - \frac{n}{b}K_n. 
\end{equation}
We have for the pressure:
\begin{eqnarray*}
\fl \frac{dP}{dr} &= D \frac{db}{dr}\frac{d}{db}\left(\frac{K_2(b)}{b^2}\right) = 
	 D \frac{db}{dr}\left( \frac{K_2'}{b^2} - \frac{2K_2}{b^3}\right) = D\frac{db}{dr}\left( -\frac{K_1}{b^2} - \frac{4K_2}{b^3}\right) \\
\fl	 &= -\frac{1}{b}D\frac{K_2}{b}\left(\frac{K_1}{K_2} + \frac{4}{b} \right)\frac{db}{dr} 
	 =-\frac{P+\rho c^2}{b}\frac{db}{dr}
\end{eqnarray*}
Hence, Tolman relation (\ref{eq:tolman}) holds. q.e.d.

So, in General Relativity, chemical potential is normally including the rest mass \cite{Klein:1949}. The condition $\alpha = const.$ will also be verified numerically as in Figure \ref{fig:delta_a}, validating the self-consistency of our approach.

\subsection{Boundary condition}

	The system of equations (\ref{eq:TOV}), (\ref{eq:mass_d}), (\ref{eq:tolman}) for the equation of state (\ref{eq:eos}) does not have any solution with vanishing density (and hence pressure) at finite edge for any initial conditions. Equivalently, if density is spread indefinitely the entropy increases and no critical point $\delta S = 0$ can be reached. This is true in the Newtonian limit, as well as in General Relativity for an ideal gas. However, in practise relaxation to equilibrium occurs only in a finite region of physical space and statistical theory must be applied in this subdomain, while on the outside different conditions occur. This is common in astrophysical systems, e.g. the hot x-ray emitting intracluster medium of galaxy clusters, which is nearly isothermal inside the virial radius, globular clusters that are isothermal, though of finite size due to galactic tidal forces, the core of stars, surrounded by the outer layers (leading to the well-known Sch\"onberg-Chandrashekhar limit) and the Bonnor-Ebert spheres of interstellar medium. This ``maximum entropy bubble'' may be achieved by \textit{incomplete relaxation} introduced by Lynden-Bell \cite{LyndenBell:1966bi}. It may be modelled with various ways. One approach is to truncate the momentum with a cut-off like in the so called King models \cite{King:1966fn} or more refined ones \cite{Hjorth:1991, Hjorth:1993}. Consult also \cite{Merafina:1989,Chavanis21051998, Chavanis:2002gg, BisnovatyiKogan:2006cw, Casetti:2012}. 
	
	 Another approach is to confine the system artificially inside a ``box'', i.e. truncate in the physical, and not momentum, space. This delimits the region within statistical theory applies by action of some external pressure at the boundary. Practically, it accounts for solving the system (\ref{eq:TOV}), (\ref{eq:mass_d}), (\ref{eq:tolman}), (\ref{eq:eos}) until some boundary $r=R$, where pressure and density take some nonzero finite value. This is the approach I use here. Although it may seem more unrealistic, it is the one leading to further theoretical insight and is the common method to use not only in gravitational \cite{Antonov:1962,Bell:1968,Cocke:1965,SWZ:1981,Hawking:1976,Gross:1982,Pavon:1988} but also in ordinary (neglecting self-gravity) thermodynamics, where the exact same problem arises (no equilibrium solution with finite mass at finite space and vanishing edge density). So that, although all theoretical work in ordinary thermodynamics is performed using these unphysical ``walls'', the theory remains apparently still extremely useful for atmospheric physics. The use of the box should be realized under this theoretical spirit, in the current work. Also hold in mind, that one of the purposes of this work is to generalize and unify the Antonov problem \cite{Bell:1968} with the radiation problem \cite{SWZ:1981} in a common framework, bridging the gap between these extreme limits\footnote{
	 I emphasize that no new result is given regarding the specific Antonov and radiation problems, apart from the understanding that they are just the two limiting cases of infinite other possibilities regarding temperature values and number of particles for an ideal gas, that lead to similar type of instabilities.}. 
	 Both problems are formulated using confinement in a box. Additionally, the application of section \ref{sec:NS} on neutron stars is relevant to the box set-up, since the isothermal core is sustained by the external pressure of outer layers.
	 
	 Let me comment on differences between the box model and the truncated -King type- model. In the box model 
one would expect the external pressure to enable the gas sustain arbitrarily high energy particles. On the contrary, in the truncation model only low energy particles are sustained, so that the main contribution to energy will come from the rest mass with positive thermal energy and negative gravitational energy being probably approximately counterbalanced in this case.	Therefore, in the box model much higher total mass-energy is expected to be able to be sustained, as is also noted in section \ref{sec:NS_micro} below the equation (\ref{eq:Mmax_micro}).
	 
	 One may still worry on the matching with the ``walls'' and the outside region in the current box model. However, this does not affect the results for the interior in the current work, and may be treated only during the construction of specific models of physical situations under specific assumptions. Regarding section \ref{sec:NS}, that is an application to neutron stars, the matching problem is not touched upon, because the purpose is not to construct an exact model of neutron stars, but only to estimate the maximum mass. The result is affected only by the edge density value, as we will see, and not by any other detail of the matching between the core (box) and the crust (outside region). Note also that this matching has been extensively studied in the past \cite{Haensel:2007}.

\section{Gravothermal Instabilities in General Relativity}\label{sec:GI}

The term `gravothermal' in the present context refers to the self-gravitating classical ideal gas, for which only thermal and gravitational pressure are present. Thermal pressure is the only one that can halt gravitational collapse in this case. The thermal component, however, contributes also to the gravitational component, since it gravitates as well. This fact captures the essence of the Tolman-Ehrenfest effect \cite{Tolman:1930,Tolman-Ehrenfest:1930,Tolman:1939}. At thermal equilibrium, there appears a temperature gradient that counterbalances the additional gravitational component of heat. However, as we will see here, there is a critical point beyond which gravitational component of heat overcomes its stabilizing outward pointing thermal pressure and the system becomes unstable. This is the \textit{high energy gravothermal instability} presented here. 

In addition the term `gravothermal' refers to the `gravothermal catastrophe' \cite{Bell:1968}, that is the instability of the self-gravitating ideal gas in the Newtonian limit. In this case, the system collapses because the thermal energy is so much diminished that can no longer halt gravitational collapse. The relativistic generalization of this phenomenon will also be presented here. This is the \textit{low energy gravothermal instability}.

As discussed previously in section \ref{sec:TE}, the TOV equation (\ref{eq:TOV}), the mass equation (\ref{eq:mass_d}), the Tolman relation (\ref{eq:tolman}) along with the equation of state (\ref{eq:eos}) describe a self-gravitating classical ideal gas in static thermal equilibrium in General Relativity and form the system of equations we have to solve. As we have already noted, there is no solution with vanishing edge pressure for an ideal gas and therefore the gas has to be confined by some external pressure at the edge we denote $R$. Thus, the system 
(\ref{eq:TOV}), (\ref{eq:mass_d}), (\ref{eq:tolman}), (\ref{eq:eos}) will be solved for initial conditions $\rho(0) = \rho_0$, $b(0) = b_0$, $\hat{M}(0) = 0$ until some truncated $r=R$.

In \ref{app:A} is shown that this system of equations gives the Emden equation in the Newtonian limit, i.e. the Poisson equation in spherical coordinates for a classical ideal gas. It has been argued in the past that the TOV equation with a linear equation of state $p = q \rho c^2$, $q = const.$, is the relativistic generalization of Emden equation, since it reduces to Emden for $q\rightarrow 0$. However, the linear equation of state is not the equation of state of the classical ideal gas. Hence, I argue, that the more proper point of view is that this system (\ref{eq:TOV}), (\ref{eq:mass_d}), (\ref{eq:tolman}), (\ref{eq:eos}), given here, is the relativistic generalization of Emden equation.

We have to generate series of thermal equilibria for fixed number of particles, i.e. fixed rest mass $\mathcal{M}$. 
An extremum of total energy $M$ designates a turning point of stability, i.e.\ the point at which an instability 
sets in, under conditions of constant energy (microcanonical ensemble), while an extremum of Tolman temperature 
designates a turning point of stability under conditions of constant Tolman temperature, i.e.\ in the presence of a heat bath. This follows from Poincar\'e theorem of linear series of equilibria \cite{Poincare,Bell:1968,Katz:1978}.

In order to work numerically, the following dimensionless variables are defined:
\begin{equation}
\fl	x = r\sqrt{4\pi G \rho_0\frac{b_0}{c^2}} \; ,\; u = \hat{M}G\frac{b_0}{c^2}\sqrt{4\pi G \rho_0\frac{b_0}{c^2}}
	\; ,\; \rho = \rho_0 e^{-y} \; ,\; b = b_0 e^{-\psi}.
\end{equation}
The system (\ref{eq:TOV}), (\ref{eq:mass_d}), (\ref{eq:tolman}), (\ref{eq:eos}) becomes:
\begin{eqnarray}
\fl
\label{eq:TOV_ND}	\frac{d y}{dx} &= \left(1+\mathcal{F}(b) - \frac{\frac{d\mathcal{F}(b)}{db}}{1+\mathcal{F}(b)}\right)
	\left(\frac{u}{x^2}e^{-\psi} + \frac{xe^{-y}}{b_0(1+\mathcal{F}(b))}\right)\left( 1 - \frac{2u}{b_0 x} \right)^{-1} 
\\
\fl
\label{eq:tolman_ND}		\frac{d\psi}{dx} &= -\frac{e^{\psi}}{b_0} 	\left(\frac{u}{x^2}e^{-\psi} + \frac{xe^{-y}}{b_0(1+\mathcal{F}(b))}\right)\left( 1 - \frac{2u}{b_0 x} \right)^{-1}
\\
\fl
\label{eq:mass_ND}	\frac{d u}{dx} &= x^2 e^{-y}	
\end{eqnarray}
with initial conditions $y(0) = \psi(0) = u(0)= 0$. The system is integrated until some point:
\begin{equation}
	z = R\sqrt{4\pi G \rho_0\frac{b_0}{c^2}}.
\end{equation}
The derivative of $\mathcal{F}$ is calculated using the recursive formulas:
\begin{eqnarray*}
	\frac{dK_n}{db} &= -K_{n+1} + \frac{n}{b}K_n \\
	&= -K_{n-1} - \frac{n}{b}K_n. 
\end{eqnarray*}
It is equal to:
\begin{equation}
	\frac{d \mathcal{F}(b)}{db} = \left( \frac{K_1}{K_2}\right)^2 + \frac{3}{b}\frac{K_1}{K_2} - \frac{3}{b^2} - 1
\end{equation}

\begin{figure}[ht]
\begin{center}
	\subfigure[$\mathcal{M} = 0.01 M_S$]{ \label{fig:ksi_1e-2_Tolman}\includegraphics[scale = 0.4]{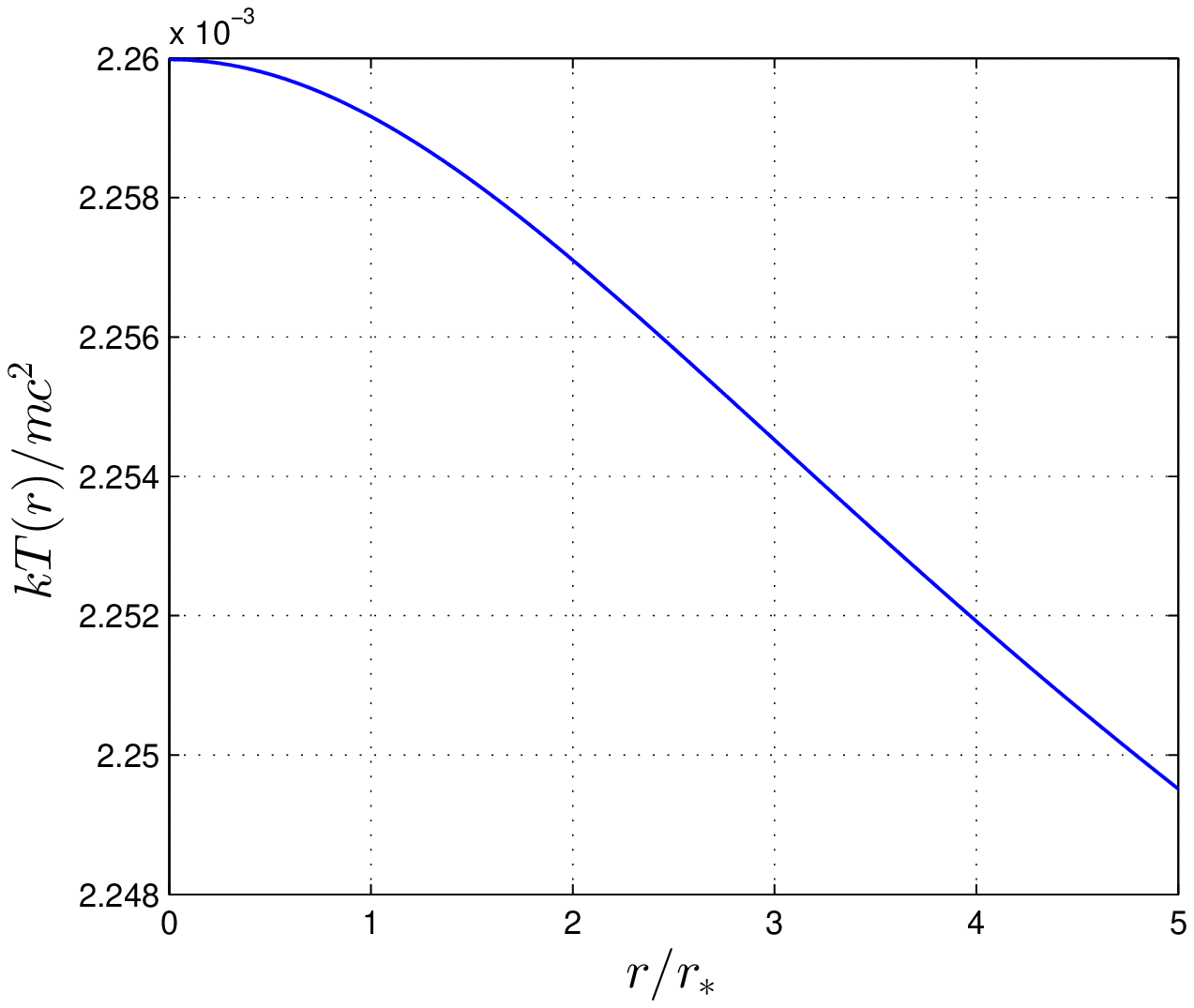} } 
	\subfigure[$\mathcal{M} = 0.25 M_S$]{ \label{fig:ksi_25e-2_Tolman}\includegraphics[scale = 0.4]{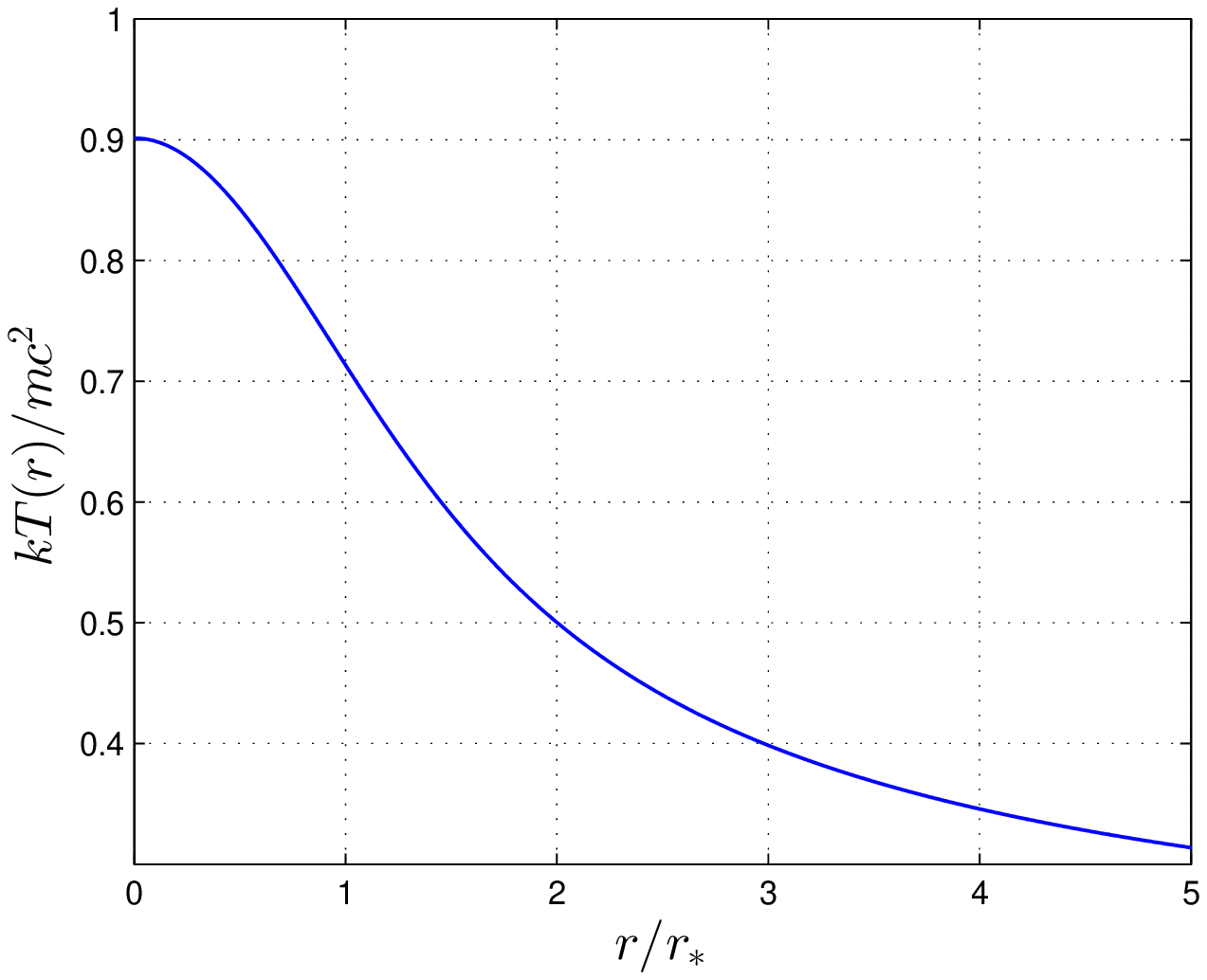} } 
	\caption{The Tolman-Ehrenfest effect for the temperature. The proper temperature versus the radius for two solutions. One with small total rest mass, namely $2G\mathcal{M}/Rc^2 = 0.01$, in (a) and one with big total rest mass, namely $2G\mathcal{M}/Rc^2 = 0.25$, in (b). The values of the constants appearing in the plots are $r_* = 1/\sqrt{4\pi G\rho_0 m\beta_0}$ and $M_S = Rc^2/2G$.
	\label{fig:Tolman_Ehrenfest}}
\end{center} 
\end{figure}

\begin{figure}[ht]
\begin{center}
	\subfigure[$\mathcal{M} = 0.01 M_S$]{ \label{fig:ksi_1e-2_Chemical}\includegraphics[scale = 0.4]{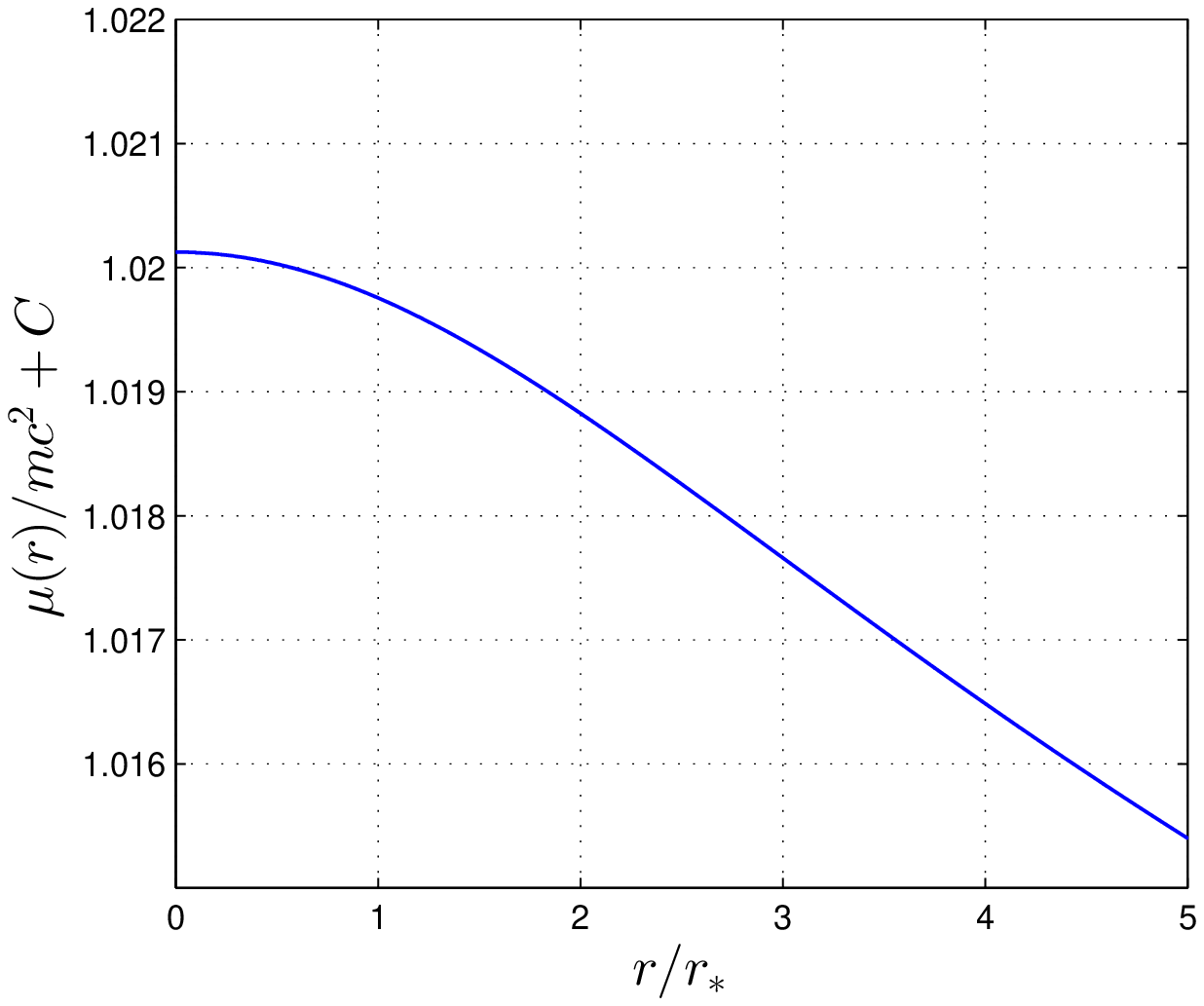} } 
	\subfigure[$\mathcal{M} = 0.25 M_S$]{ \label{fig:ksi_25e-2_Chemical}\includegraphics[scale = 0.4]{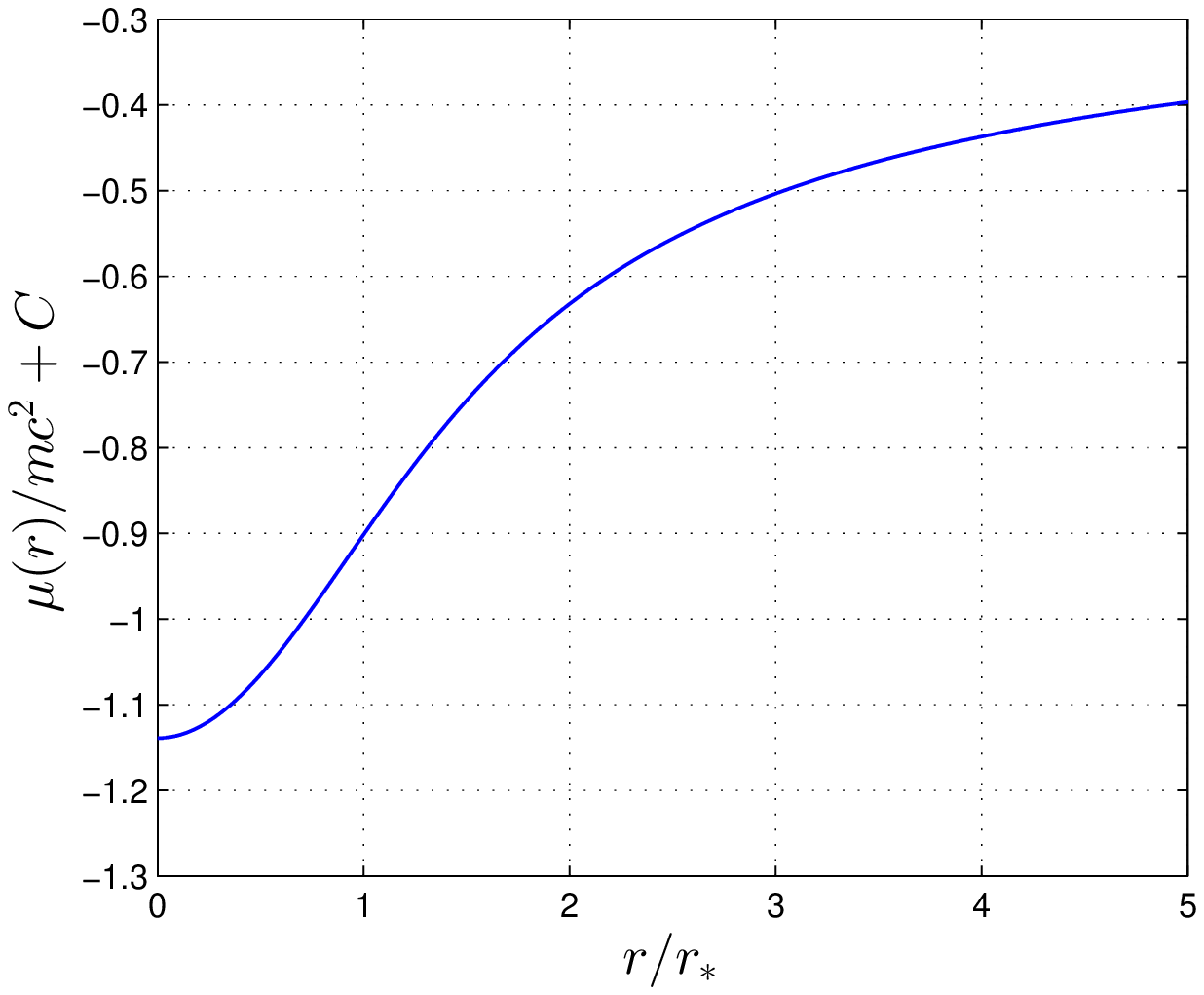} } 
	\caption{The Tolman-Ehrenfest effect for the chemical potential. The proper chemical potential versus the radius for two solutions. One with small total rest mass, namely $2G\mathcal{M}/Rc^2 = 0.01$, in (a) and one with big rest mass, namely $2G\mathcal{M}/Rc^2 = 0.25$, in (b). The values of the constants appearing in the plots are $C = \ln(8\pi m^4 c^3/\rho_0 h^3)$, $r_* = 1/\sqrt{4\pi G\rho_0 m\beta_0}$ and $M_S = Rc^2/2G$.
	\label{fig:Chemical_Potential}}
\end{center} 
\end{figure}

In Figures \ref{fig:Tolman_Ehrenfest} and \ref{fig:Chemical_Potential}, the Tolman-Ehrenfest effect, as expressed by equations (\ref{eq:Tolman}) and (\ref{eq:mu_Tolman}), is demonstrated. The temperature and the chemical potential versus radius $r$ are plotted for two solutions, one with small rest mass and one with big rest mass. For small rest masses over $R$ the effect is small, while as the rest mass over $R$ increases it becomes important. The chemical potential may be positive as discussed in previous section. This may occur at low temperatures. 

\begin{figure}[ht]
\begin{center}
	\subfigure[$\mathcal{M} = 0.01 M_S$]{ \label{fig:ksi_1e-2_delta_a}\includegraphics[scale = 0.4]{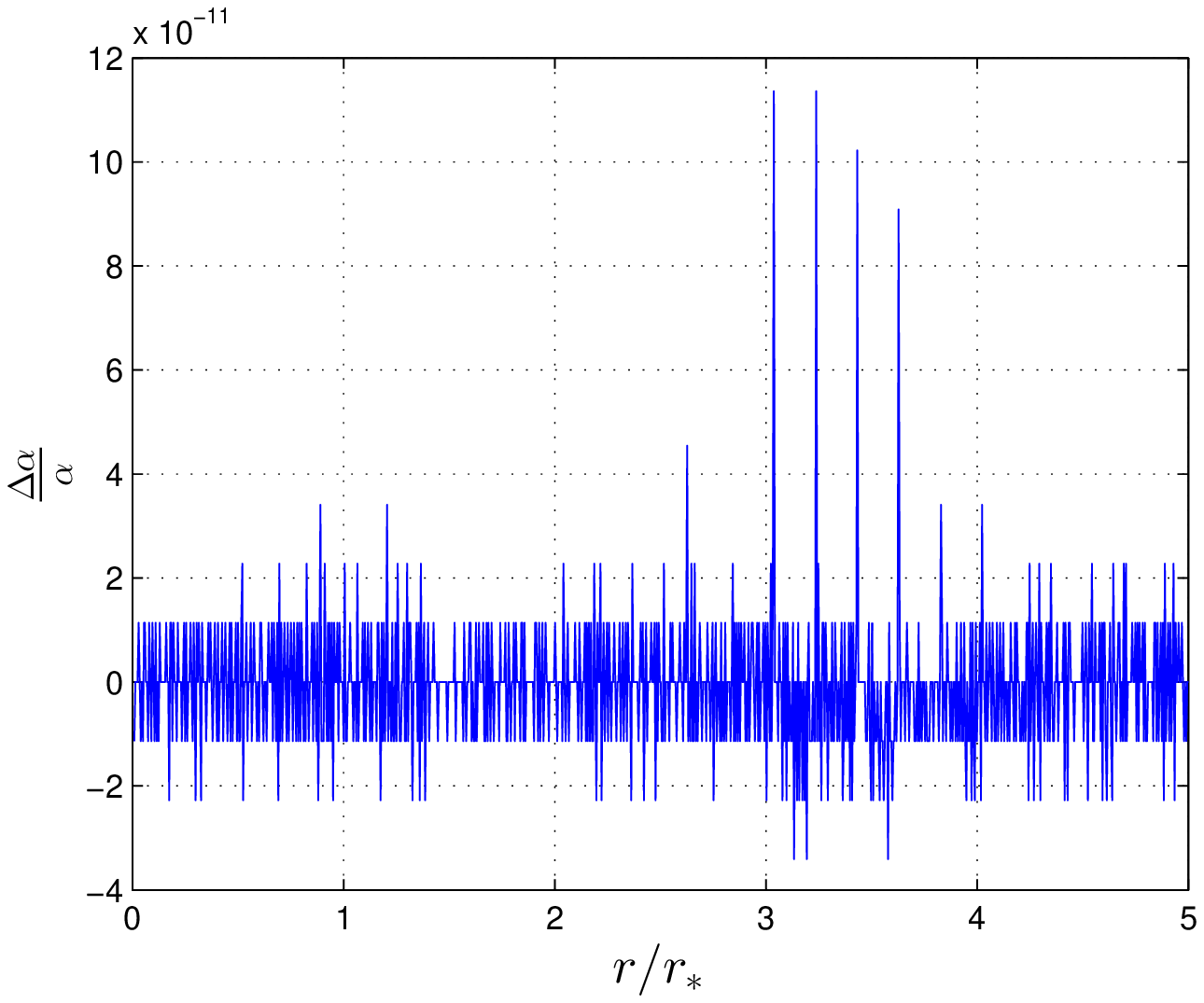} } 
	\subfigure[$\mathcal{M} = 0.25 M_S$]{ \label{fig:ksi_25e-2_delta_a}\includegraphics[scale = 0.4]{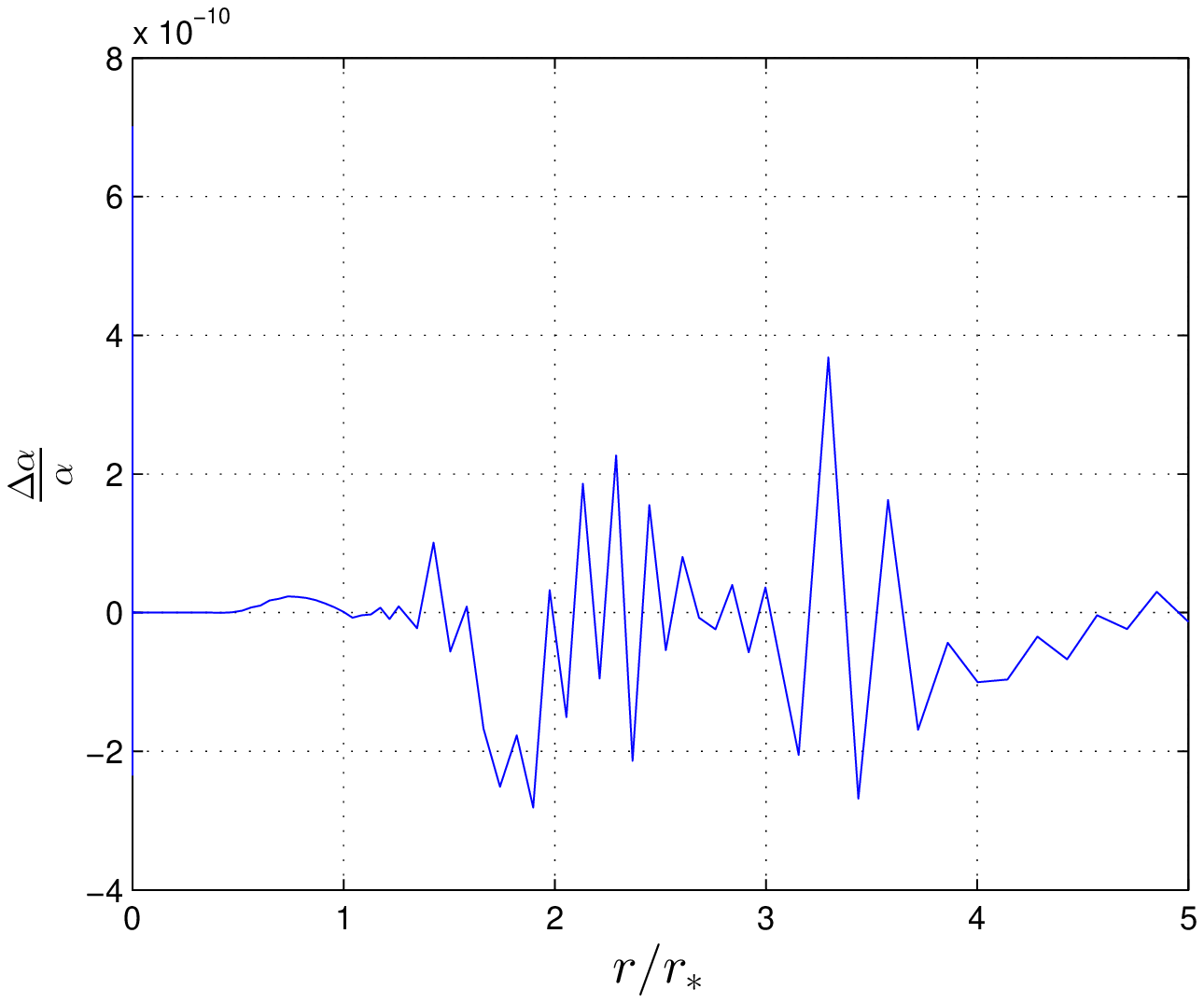} } 
	\caption{The relative change of $\alpha = \beta\mu$ for two thermal equilibria (a) and (b). It is verified the constancy of $\alpha$ within the relative tolerance I used, namely $10^{-8}$, for the solution of the differential system. This verifies that the particle rest mass should be included in the definition of chemical potential in General Relativity, as discussed in section \ref{sec:EOS}, and most importantly verifies the self-consistency of our framework. The values of the constants appearing in the plots are $r_* = 1/\sqrt{4\pi G\rho_0 m\beta_0}$ and $M_S = Rc^2/2G$, while $\mathcal{M}$ is the total rest mass.
	\label{fig:delta_a}}
\end{center} 
\end{figure}

In Figure \ref{fig:delta_a} is plotted the relative change of $\alpha = \beta \mu$ for two solutions. It is verified the constancy of $\alpha$ as required by thermal equilibrium. This verifies the self-consistency of the framework and also shows that in the definition of chemical potential, the particle rest mass $m$ should be included in General Relativity, as discussed in previous section.

The total rest mass $\mathcal{M}$ of the solution depends on the pair $(b_0,z)$. In order to generate series of equilibria with the same rest mass, let define:
\begin{equation}\label{eq:ksi}
		\xi = \frac{2G\mathcal{M}}{Rc^2}.
\end{equation}

\begin{figure}[ht]
\begin{center}
	\includegraphics[scale = 0.55]{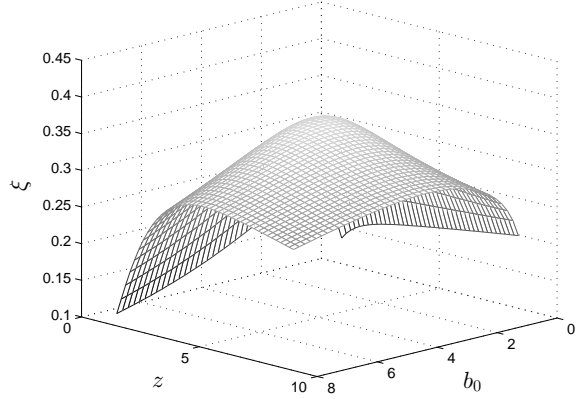}  	
	\caption{The rest mass over radius $\xi = 2G\mathcal{M}/Rc^2$ with respect to the dimensionless quantity $z = R\sqrt{4\pi G \rho_0 b_0/c^2}$ and the dimensionless central inverse temperature $b_0 = mc^2/kT_0$. There is evident a global maximum $\xi_{max} = 0.35$ at $(b_0,z) = (3.18,3.97)$.
	\label{fig:Mrest3D}}
\end{center} 
\end{figure}

The quantity $\xi$ with respect to $(z,b_0)$ is plotted in Figure \ref{fig:Mrest3D}. It presents a maximum at:
\begin{equation}
	\mathcal{M}_{max} = 0.35 M_S
\end{equation}
where $M_S = Rc^2/2G$ is the Schwarzschild mass. This value signifies the maximum number of non-interacting particles a sphere of radius $R$ can hold without collapsing, under any conditions. At this maximum correspond the following values:
\begin{equation}
\fl	\frac{2GM}{Rc^2} = 0.41 \; ,\; \frac{ER}{G\mathcal{M}^2} = 0.96 \; ,\; \frac{kT_0}{mc^2} = 0.31 \; , \; \frac{k\tilde{T}}{mc^2} = 0.19 \; , \; z = 3.97
	\; ,\; \frac{\rho_0}{\rho_R} = 27
\end{equation}

Each point of the surface in Figure \ref{fig:Mrest3D} represents a thermal equilibrium. In order to generate series of equilibria with the same rest mass, I developed a computer program that solves the system (\ref{eq:TOV_ND}-\ref{eq:mass_ND}) for various values $(b_0,z)$ keeping $\xi$ fixed. Solving for various $\xi$ may be regarded as solving for various $\mathcal{M}$ for fixed radius. The series of equilibria are defined by the intersection of a plane perpendicular to $\xi$-axis with $\xi$.

\begin{figure}[ht]
\begin{center}
	\includegraphics[scale = 0.55]{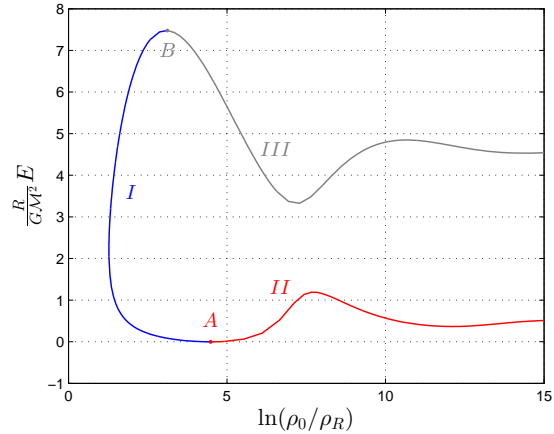}  	
	\caption{The series of equilibria for $\xi = 0.25$. The gravothermal energy $E$ of each equilibrium is plotted for fixed edge radius $R$ and total rest mass $\mathcal{M}$ with respect to the density contrast $\rho_0/\rho_R$. At point $A$ sets in the low energy gravothermal instability (relativistic generalization of Antonov instability) and at point $B$ sets in the high energy gravothermal instability in the microcanonical ensemble. No equilibria exist for $E<E_A$ and $E>E_B$ and in addition the branches $II$ and $III$ are unstable in the microcanonical ensemble designating also the critical values of density contrast at which microcanonical gravothermal instabilities set in.
	\label{fig:EvsDC}}
\end{center} 
\end{figure}

Let $E$ denote the gravothermal energy (a form of \textit{binding energy}), that is:
\begin{equation}
	E = (M - \mathcal{M}) c^2
\end{equation}
In Figure \ref{fig:EvsDC} is demonstrated the series of equilibria for $\xi = 0.25$. The gravothermal energy of each equilibrium is plotted assuming fixed radius $R$ and total rest mass $\mathcal{M}$ with respect to the density contrast $\rho_0/\rho_R$. The low energy gravothermal instability sets in at the equilibrium $A$ (generalization of Antonov instability), while the high energy gravothermal instability sets in at equilibrium $B$. This means not only that no equilibria  exist (with rest mass $\xi = 0.25$ for fixed radius) for $E<E_A$ and $E>E_B$ but also that the branches $II$ and $III$ are unstable under conditions of constant total energy (microcanonical ensemble). So that, points $A$ and $B$ designate also  critical values of density contrast at which gravothermal instabilities in the microcanonical ensemble set in. 

\begin{figure}[ht]
\begin{center}
	\subfigure[At $\xi \rightarrow 0$, the series of equilibria converge to the dust case at low temperatures.]{ \label{fig:VariousE_low_inst}\includegraphics[scale = 0.4]{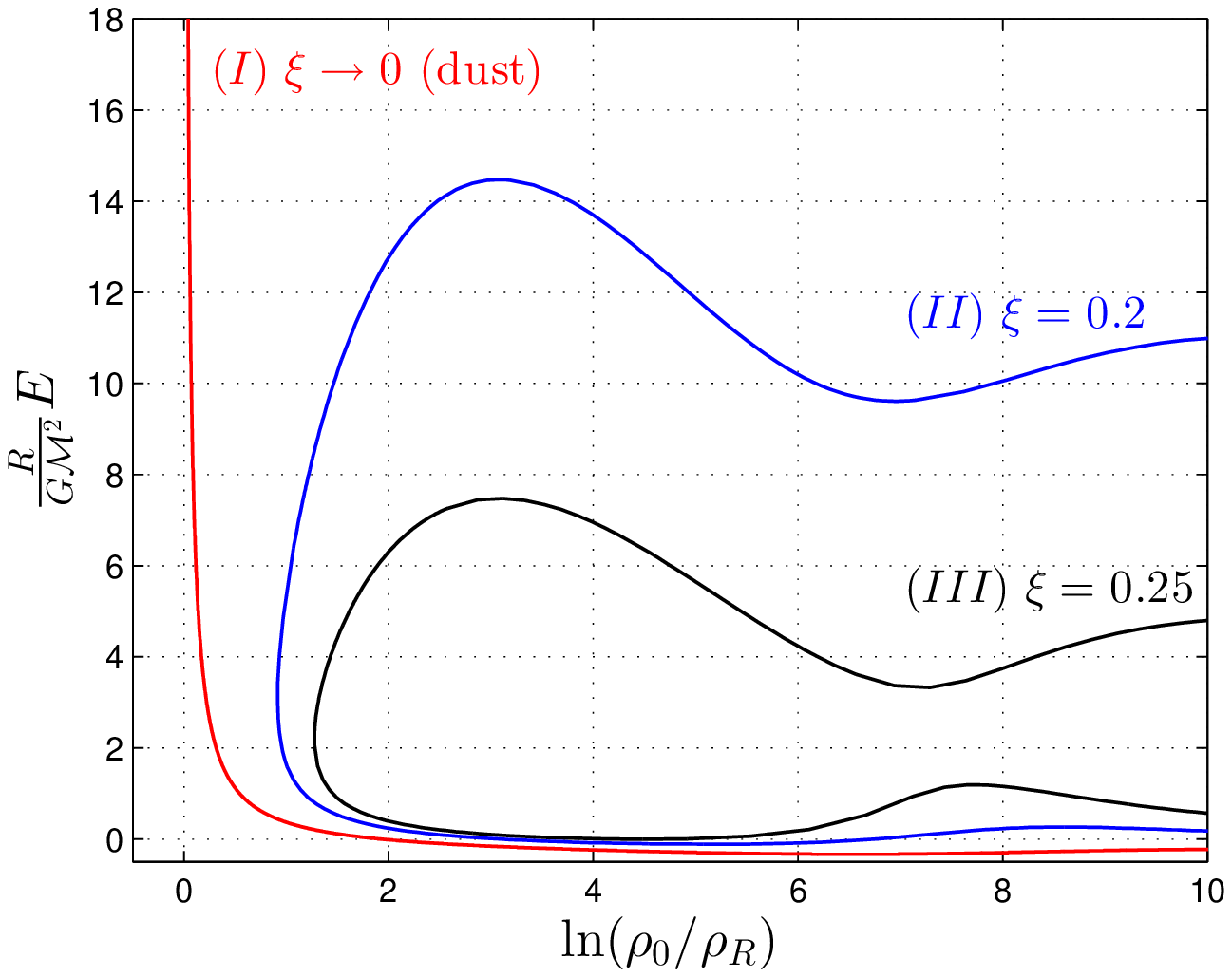} } 
	\subfigure[At $\xi \rightarrow 0$, the series of equilibria converge to the radiation case at high temperatures.]{ \label{fig:VariousE_high_inst}\includegraphics[scale = 0.4]{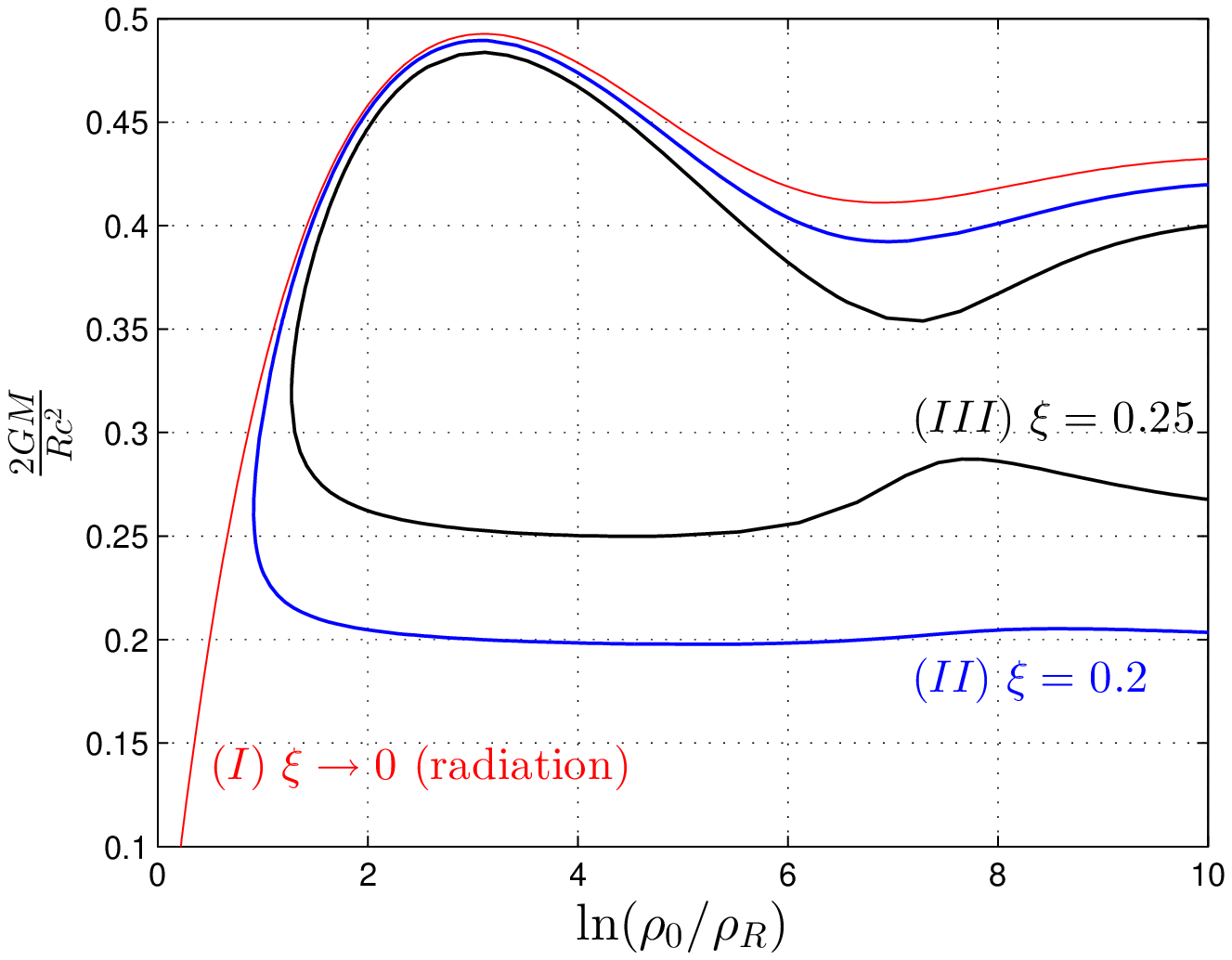} } 
	\caption{The series of equilibria for various $\xi$, namely series $(I)$ for $\xi \rightarrow 0$, series $(II)$ for $\xi = 0.2$ and series $(III)$ for $\xi = 0.25$. In (a) is plotted the gravothermal energy $E$, while in (b) the total mass $M$, for fixed edge radius and total rest mass with respect to the density contrast. In (a), series (I) corresponds to the dust (Newtonian) limit which is recovered for low temperatures $k\tilde{T} \ll mc^2$. In (b), series (I) corresponds to the radiation (Ultra-relativistic) limit which is recovered for high temperatures $k\tilde{T} \gg mc^2$.
	\label{fig:VariousE}}
\end{center} 
\end{figure}

\begin{figure}[ht]
\begin{center}
	\subfigure[]{ \label{fig:TvsDC}\includegraphics[scale = 0.4]{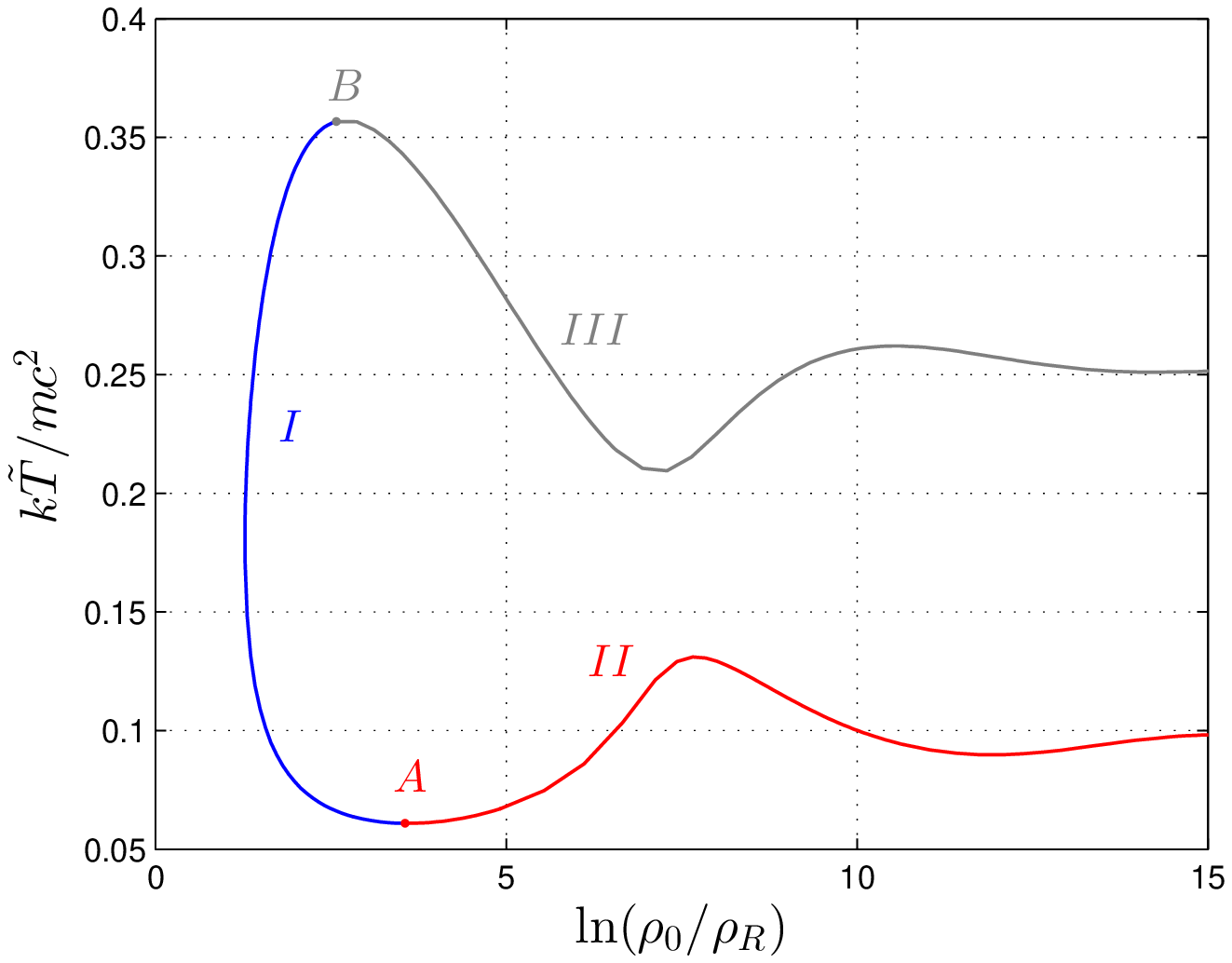} } 
	\subfigure[]{ \label{fig:VariousT_low_inst}\includegraphics[scale = 0.4]{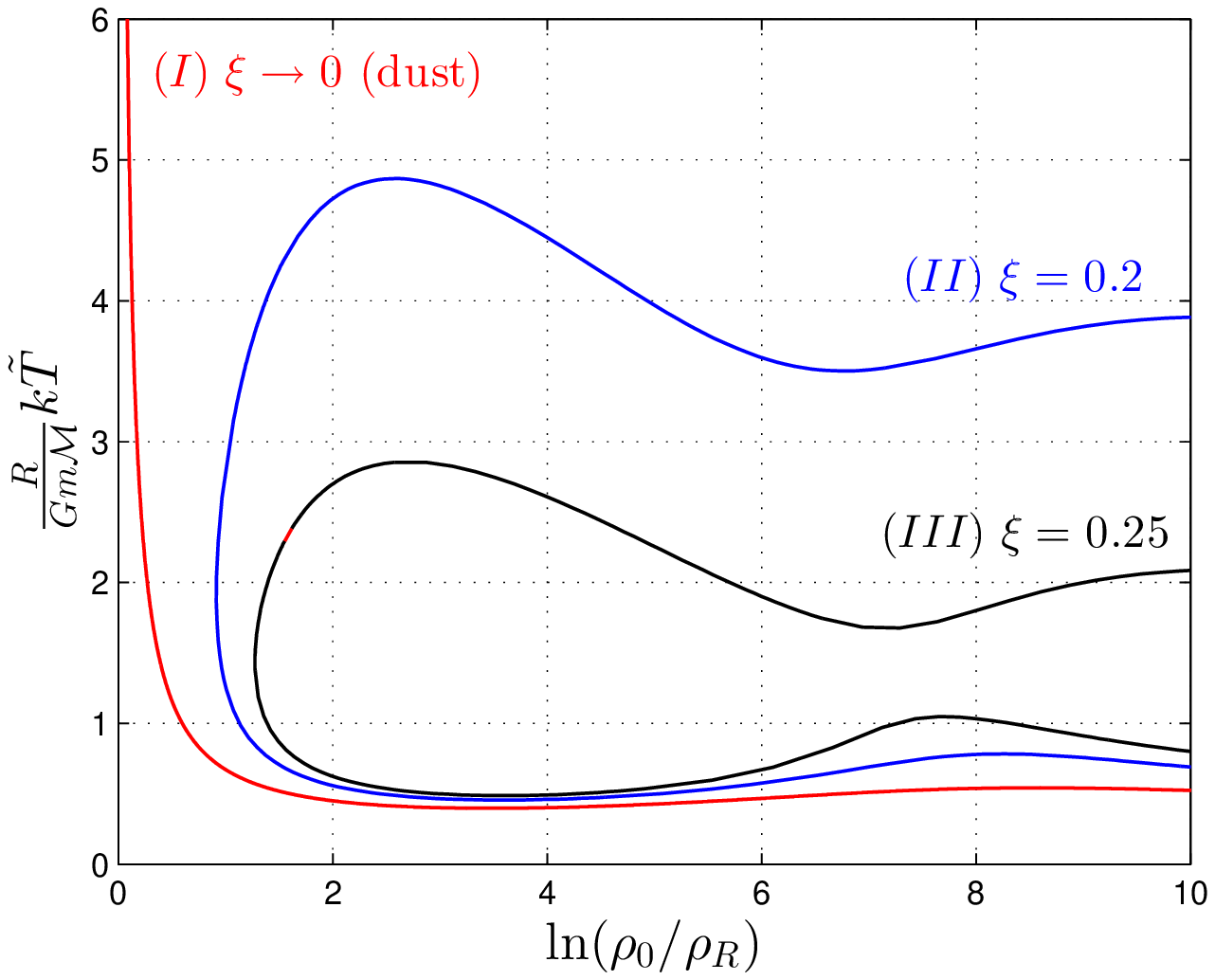} } 
	\caption{In (a) is plotted the Tolman temperature $\tilde{T}$ for $\xi = 0.25$ and fixed edge radius and total rest mass with respect to the density contrast. At point $A$ sets in the low energy gravothermal instability (relativistic generalization of Newtonian isothermal collapse) and at point $B$ sets in the high energy gravothermal instability in the canonical ensemble. No equilibria exist for $\tilde{T}<\tilde{T}_A$ and $\tilde{T}>\tilde{T}_B$ and in addition the branches $II$ and $III$ are unstable in the canonical ensemble designating also the critical values of density contrast at which canonical gravothermal instabilities set in. In (b) is plotted the Tolman temperature with respect to the density contrast for various $\xi$ values.
	\label{fig:T}}
\end{center} 
\end{figure}

I stress that \textit{both low and high energy microcanonical gravothermal instabilities set in as the specific heat goes from negative (stable equilibria) to positive (unstable equilibria) values}, just like Antonov instability.

In Figure \ref{fig:VariousE} is plotted the energy with respect to the density contrast for various values of total rest mass, parametrized by $\xi$. It is shown that at $\xi\rightarrow 0$ the series of equilibria converges to the dust (Newtonian) series at low temperatures and to the radiation $P = \rho/3$ series at high temperatures, as it should be according to equations (\ref{eq:eos_limitN}), (\ref{eq:eos_limitR}).

In Figure \ref{fig:TvsDC} is plotted the Tolman temperature of each equilibrium with fixed $\xi = 0.25$ assuming fixed radius $R$ and total rest mass $\mathcal{M}$ with respect to the density contrast $\rho_0/\rho_R$. The low energy gravothermal instability sets in at the equilibrium $A$ (generalization of Newtonian isothermal collapse), while the high energy gravothermal instability sets in at equilibrium $B$. This means not only that no equilibria  exist (with rest mass $\xi = 0.25$ for fixed radius) for $\tilde{T}<\tilde{T}_A$ and $\tilde{T}>\tilde{T}_B$ but also that the branches $II$ and $III$ are unstable under conditions of constant Tolman temperature (canonical ensemble). So that, points $A$ and $B$ designate also  critical values of density contrast at which gravothermal instabilities in the canonical ensemble set in.

\begin{figure}[ht]
\begin{center}
	\includegraphics[scale = 0.55]{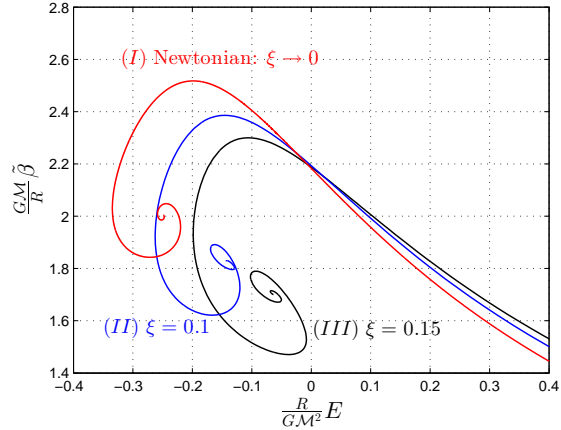}  	
	\caption{The low energy relativistic gravothermal instability (relativistic generalization of Antonov instability). The inverse Tolman temperature versus the gravothermal energy $\tilde{\beta}(E)$ is plotted for fixed radius and total rest mass, in the Newtonian limit and for exact relativistic calculations for $\xi = 0.1$ and $\xi = 0.15$. As the relativistic control parameter $\xi$ increases, both the minimum energy (gravothermal catastrophe) and the minimum temperature (isothermal collapse) increase ($\tilde{\beta}_{max}$ decreases). The Newtonian limit, given by spiral $(I)$ coincides with the, rather famous, spiral of Lynden-Bell \& Wood \cite{Bell:1968}.
	\label{fig:Spiral_upper}}
\end{center} 
\end{figure}

\begin{figure}[ht]
\begin{center}
	\includegraphics[scale = 0.55]{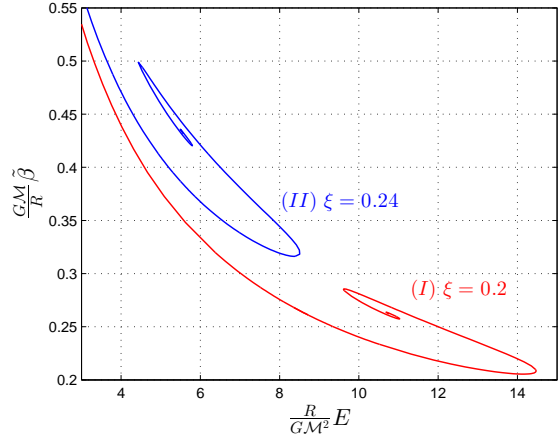}  	
	\caption{The high energy relativistic gravothermal instability. The inverse Tolman temperature versus the gravothermal energy $\tilde{\beta}(E)$ is plotted for fixed radius and total rest mass for $\xi = 0.2$ and $\xi = 0.24$. As the relativistic control parameter $\xi$ increases, both the maximum energy (microcanonical instability) and the maximum temperature (canonical instabiltiy) decrease ($\tilde{\beta}_{max}$ increases). 
	\label{fig:Spiral_lower}}
\end{center} 
\end{figure}

In Figure \ref{fig:Spiral_upper} are demonstrated the low energy gravothermal instabilities. The $\tilde{\beta}(E)$ curve is plotted, i.e the inverse Tolman temperature versus the gravothermal energy, for fixed radius and total rest mass, and for various values of the relativistic control parameter $\xi$ (with $0<\xi< 0.35$). It is evident that gravothermal catastrophe (energy minima) occurs at higher energy and isothermal collapse (temperature minima, that is $\tilde{\beta}$ maxima) at higher temperature as the system becomes more relativistic (increasing $\xi$).

\begin{figure}[ht]
\begin{center}
	\includegraphics[scale = 0.55]{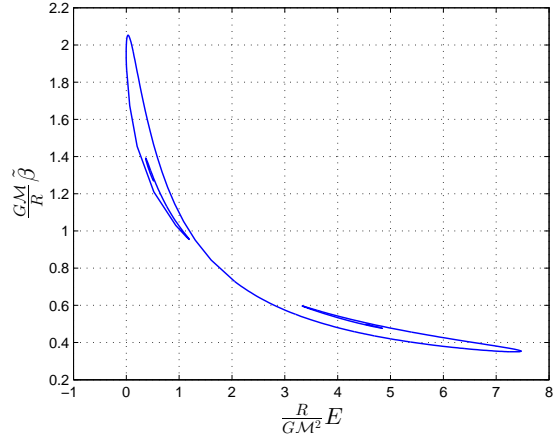}  
	\caption{Both types of gravothermal instabilities --low and high energy-- are present for \textit{any} value of $\xi < \xi_{max}$ and make themselves evident as a \textit{double spiral} in the $\tilde{\beta}(E)$ series of equilibria. This is demonstrated in this plot for $\xi = 0.25$. The Tolman inverse temperature versus the gravothermal energy is plotted for fixed radius and total rest mass using the `Newtonian' type of dimensionless variables. The upper spiral corresponds to the low energy gravothermal instability, while the lower spiral corresponds to the high energy gravothermal instability. In the microcanonical ensemble stable equilibria are only those in between the energy maxima and minima, while the rest that lie at each spiral are unstable. At every energy extremum of the spirals a new mode of instability is added. In the canonical ensemble stable equilibria are only those in between the temperature maxima and minima.
	\label{fig:DoubleSpiral_1}}
\end{center} 
\end{figure}

\begin{figure}[ht]
\begin{center}
	\includegraphics[scale = 0.55]{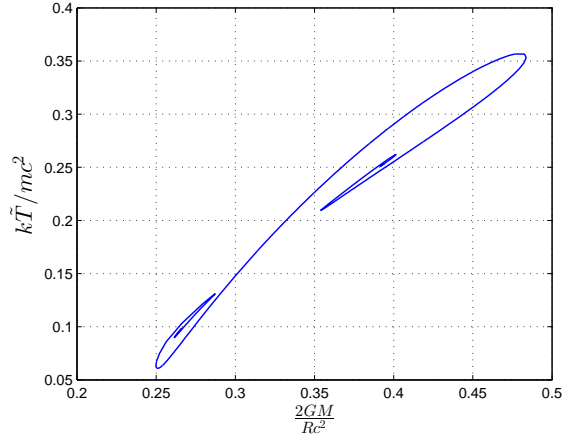} 
	\caption{The double spiral of Figure \ref{fig:DoubleSpiral_1} in `relativistic' type of dimensionless variables. The Tolman temperature versus the total energy $M$ is plotted for fixed radius and total rest mass $\mathcal{M}$. In this figure, the lower spiral corresponds to the low energy gravothermal instability, while the upper spiral corresponds to the high energy gravothermal instability. 
	\label{fig:DoubleSpiral_2}}
\end{center} 
\end{figure}

In Figure \ref{fig:Spiral_lower} are demonstrated the high energy gravothermal instabilities. The $\tilde{\beta}(E)$ curve is plotted for fixed radius and total rest mass, and for various values of the relativistic control parameter $\xi$. It is evident that both the maximum energy (microcanonical instability) and the maximum temperature (canonical instability) decrease as the system becomes more relativistic (increasing $\xi$).

For \textit{any} value of $\xi$, no matter how small, \textit{both} types --low and high energy-- of gravothermal instabilities are present. This leads to a double spiral in the $\tilde{\beta}(E)$ curve as is evident in Figures \ref{fig:DoubleSpiral_1} and \ref{fig:DoubleSpiral_2}.

\subsection{Critical quantities in the microcanonical ensemble}\label{sec:NS_micro}

Let investigate how the critical mass, radius and density contrast vary in the case of microcanonical ensemble for both low and high energy gravothermal instabilities.

\begin{figure}[ht]
\begin{center}
	\includegraphics[scale = 0.55]{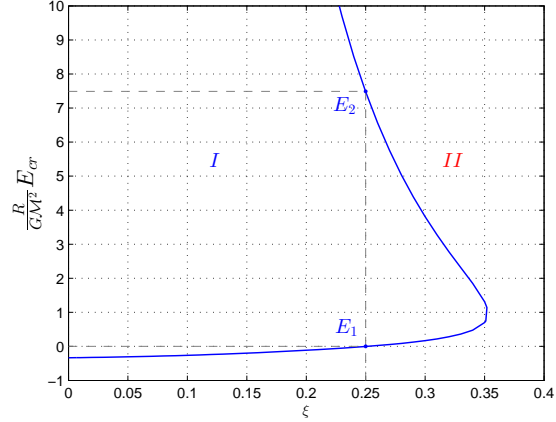} 
	\caption{The critical gravothermal energy, at which a microcanonical gravothermal instability sets in, with respect to $\xi$. Thermal equilibria exist only in the region $I$ `inside' the curve, while no equilibria exist in the `outside' region $II$. For every value of $\xi$, namely of radius and total rest mass, there are two critical energy values. The upper value $E_2(\xi)$ corresponds to the high energy relativistic gravothermal instability, while the lower value $E_1(\xi)$ to the low energy relativistic gravothermal instability. Thermal equilibria exist only in between these critical values for this fixed $\xi$. 
	\label{fig:Ecr}}
\end{center} 
\end{figure}

\begin{figure}[ht]
\begin{center}
	\subfigure[]{ \label{fig:Mcr_nose}\includegraphics[scale = 0.4]{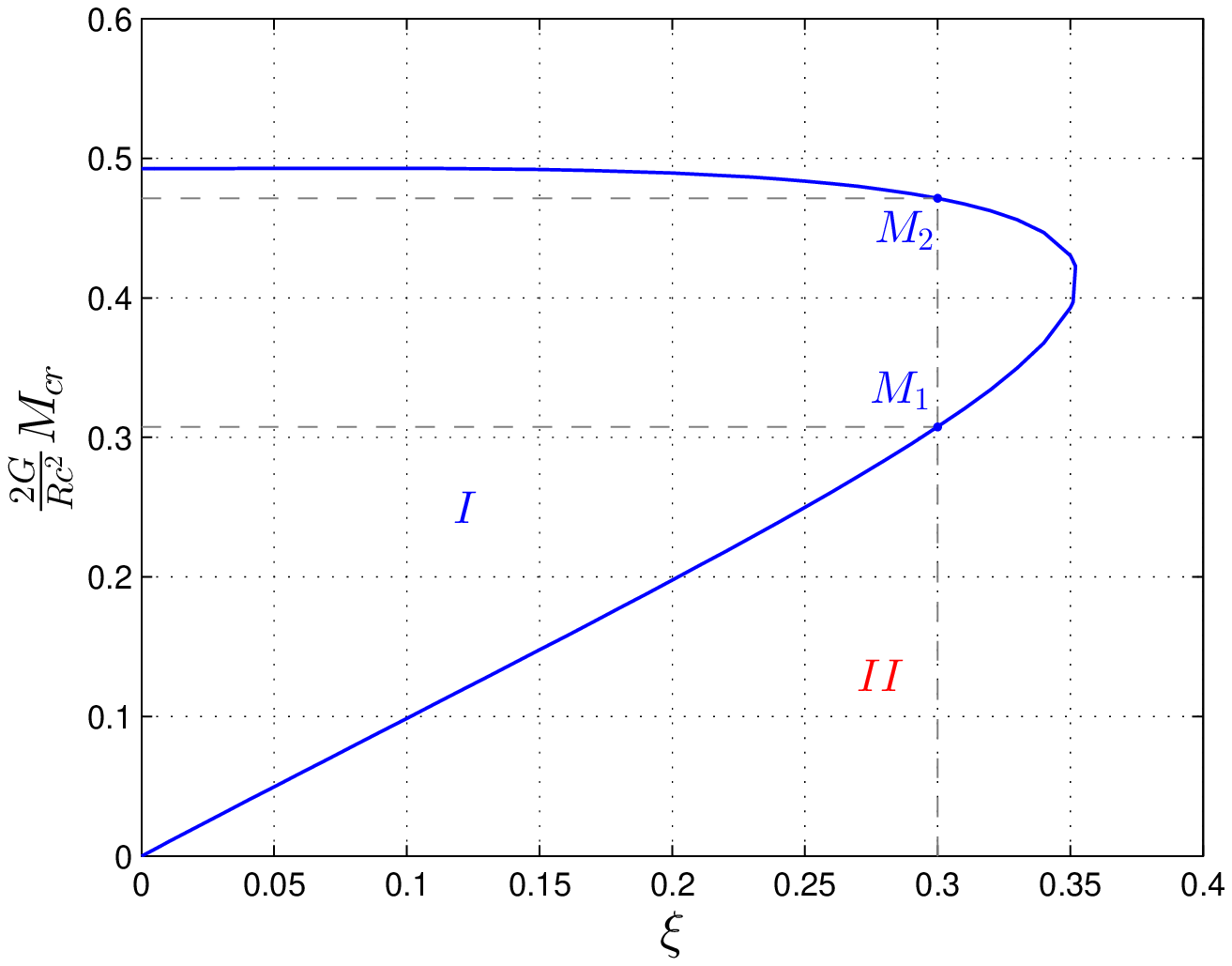} } 
	\subfigure[]{ \label{fig:Mcr_max}\includegraphics[scale = 0.4]{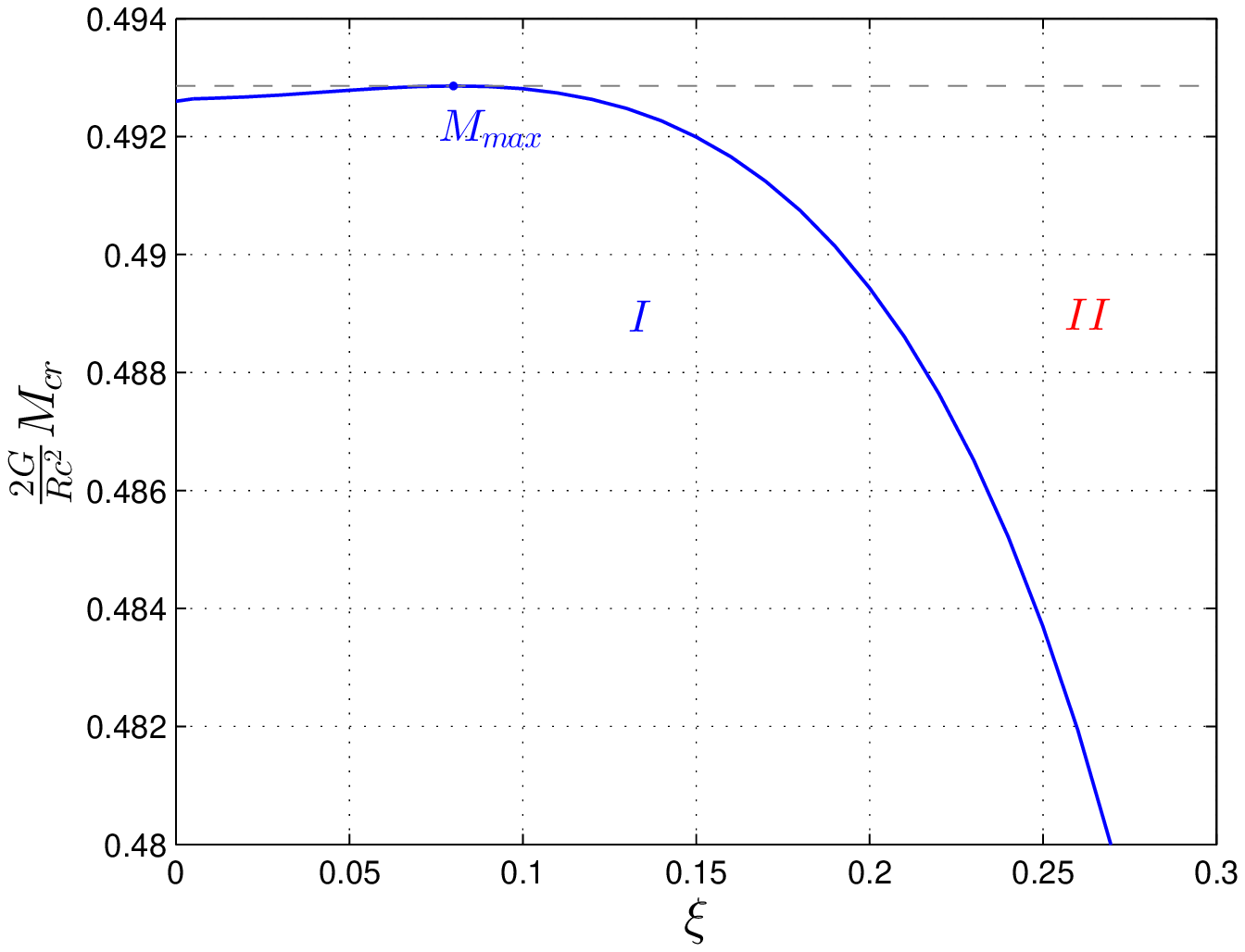} } 
	\caption{The critical total mass, at which a microcanonical gravothermal instability sets in, with respect to $\xi$. Just like Figure \ref{fig:Ecr}, for every value of $\xi$, namely of radius and total rest mass, there are two critical mass values corresponding to the low ($M_1(\xi)$) and high ($M_2(\xi)$) energy gravothermal instabilities. In (b) we focus on the high energy gravothermal instability, which is found to present an ultimate upper limit of critical mass $M_{max} = 0.493M_S$ at $\xi = 0.08$ corresponding to Tolman temperature $k\tilde{T} = 1.3mc^2$.
	\label{fig:Mcr_micro}}
\end{center} 
\end{figure}

In Figure \ref{fig:Ecr} is plotted the critical gravothermal energy with respect to $\xi$. It is evident that for every value of total rest mass and radius there are two critical energies and equilibria exist only in between these critical values. The upper value corresponds to the high energy gravothermal instability, while the lower value to the low energy gravothermal instability.

In Figure \ref{fig:Mcr_micro} is plotted the critical total mass with respect to $\xi$. From Figure \ref{fig:Mcr_max} is evident that the high energy gravothermal instability presents an ultimate upper limit of critical mass, the value of which is 
\begin{equation}\label{eq:Mmax_micro}
	M_{max} = 0.493M_S \; \mbox{, (microcanonical ensemble)}.
\end{equation}
It appears at $\xi = 0.08$ corresponding to Tolman temperature $k\tilde{T} = 1.3mc^2$. Note that this value of $\xi$ is approximately equal to the maximum total mass over radius $2GM/Rc^2 = 0.083$ in case of the momentum truncated model with no walls \cite{Merafina:1989,BisnovatyiKogan:2006cw}, while total mass of equation (\ref{eq:Mmax_micro}) is rather huge comparatively. One would suspect this is because the external pressure in the current box model enables the gas to sustain arbitrarily high energy particles, while the truncation model contains only low energy particles, so that the main contribution to energy comes from the rest mass with positive thermal energy and negative gravitational energy being counterbalanced in this case.

\begin{figure}[ht]
\begin{center}
	\subfigure[]{ \label{fig:Rcr_nose}\includegraphics[scale = 0.4]{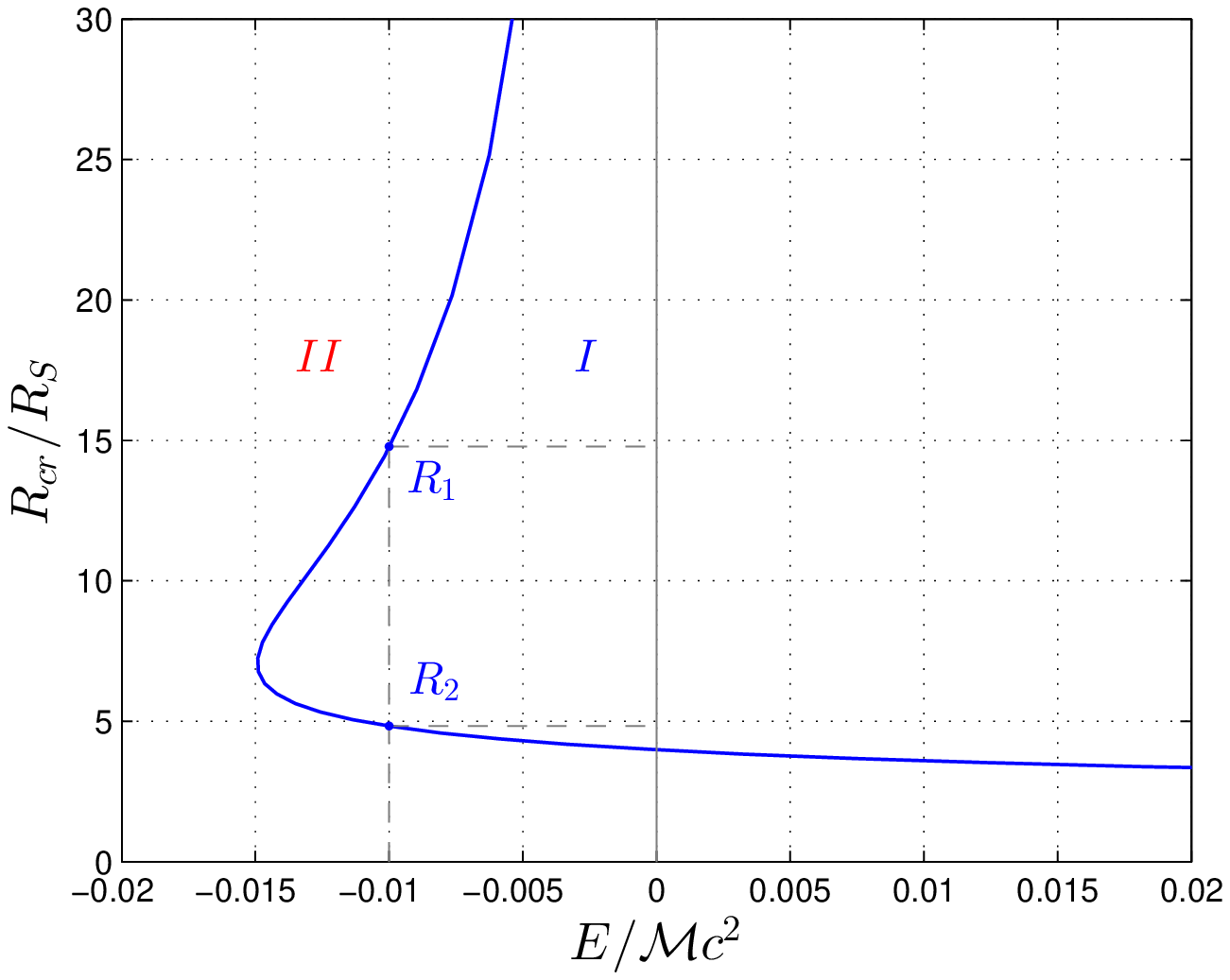} } 
	\subfigure[]{ \label{fig:Rcr_min}\includegraphics[scale = 0.4]{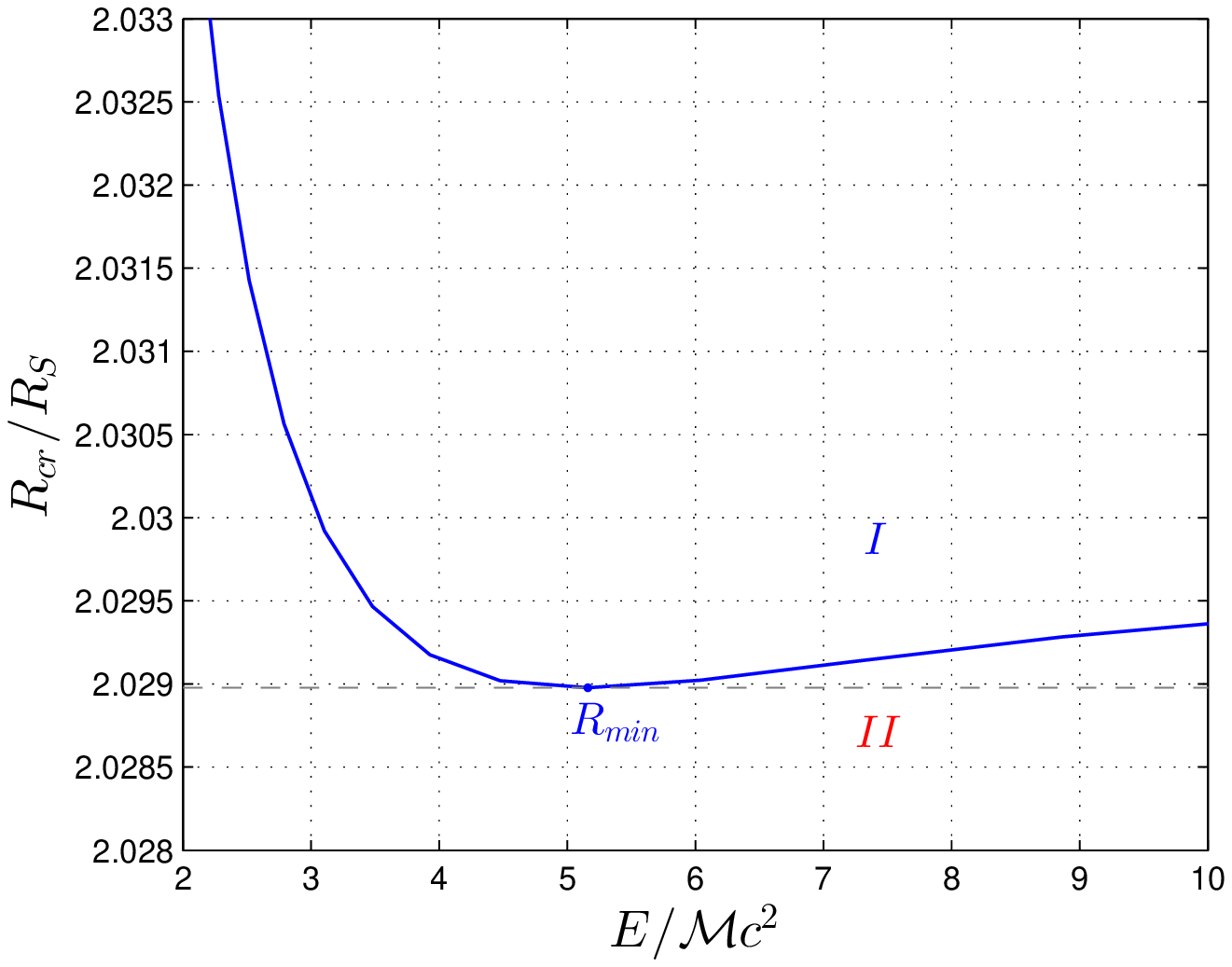} } 
	\caption{The critical radius, at which a gravothermal instability in the \textit{microcanonical ensemble} sets in, with respect to gravothermal energy for fixed total rest mass. Equilibria exist only inside region $I$ and not in the `outside' region $II$. For every fixed negative gravothermal energy there are two critical radii corresponding to the two types of gravothermal instabilities: $R_1(\xi)$ to the low energy gravothermal instability and $R_2(\xi)$ to the high energy gravothermal instability. Equilibria exist only in between these two critical values. For positive gravothermal energy only the minimum critical radius is present and equilibria exist for arbitrary large radii. In (b) is shown the ultimate minimum radius. It has the value $R_{min} = 2.03R_S$ and corresponds to the $M_{max}$ of Figure \ref{fig:Mcr_max}.
	\label{fig:Rcr_micro}}
\end{center} 
\end{figure}

In Figure \ref{fig:Rcr_micro} is plotted the critical radius in the microcanonical ensemble versus the gravothermal energy for fixed total rest mass. At each fixed \textit{negative} gravothermal energy there are two critical radii corresponding to the two types of gravothermal instabilities. The upper critical radius corresponds to the low energy gravothermal instability, while the lower critical radius to the high energy gravothermal instability. Equilibria exist only in between these two critical values. For positive gravothermal energy only the minimum critical radius is present and equilibria exist for arbitrary large radii. The ultimate minimum radius of Figure \ref{fig:Rcr_min} is
\begin{equation}
	R_{min} = 2.03R_S\; \mbox{, (microcanonical ensemble)}
\end{equation}
and corresponds to the $M_{max}$ given before.

\begin{figure}[ht]
\begin{center}
	\includegraphics[scale = 0.55]{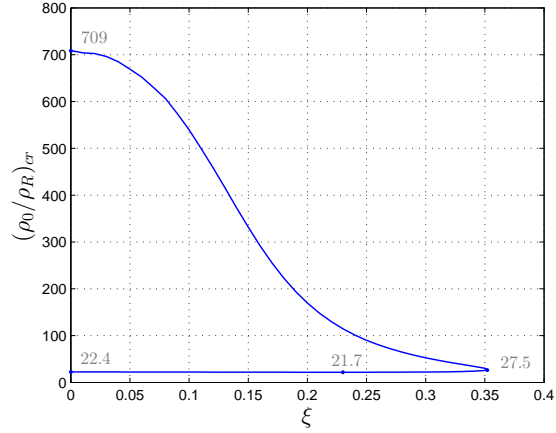}  
	\caption{The critical density contrast, i.e. the value at which a gravothermal instability sets in, in the \textit{microcanonical ensemble} versus $\xi$. Values above $27.5$, corresponding to $\xi = 0.35$, correspond to the low energy instability, while values below $27.5$ correspond to the high energy instability.
	\label{fig:DC_micro}}
\end{center} 
\end{figure}

In Figure \ref{fig:DC_micro} is plotted the critical density contrast in the microcanonical ensemble with respect to $\xi$. At each of these values a gravothermal instability in the microcanonical ensemble sets in. The critical density contrast of the low energy instability varies in the interval $[27.5,709]$, while the high energy instability in the interval $[21.7,27.5]$.

\subsection{Critical quantities in the canonical ensemble}

Relativistic instabilities in a heat bath, i.e. in the canonical ensemble, correspond to perturbations under constant Tolman temperature \cite{Roupas:2014nea}. In the Newtonian limit such an instability is called `isothermal collapse' \cite{Padman:1990,Chavanis:2001hd,Axenides:2013npb,Axenides:2013hba} and sets in at low temperatures, when the thermal pressure can no longer support gravitational attraction. Let investigate how the critical temperature, radius and density contrast vary in the case of canonical ensemble for both low and high energy gravothermal instabilities. 

\begin{figure}[ht]
\begin{center}
	\subfigure[]{ \label{fig:Tcr_can}\includegraphics[scale = 0.4]{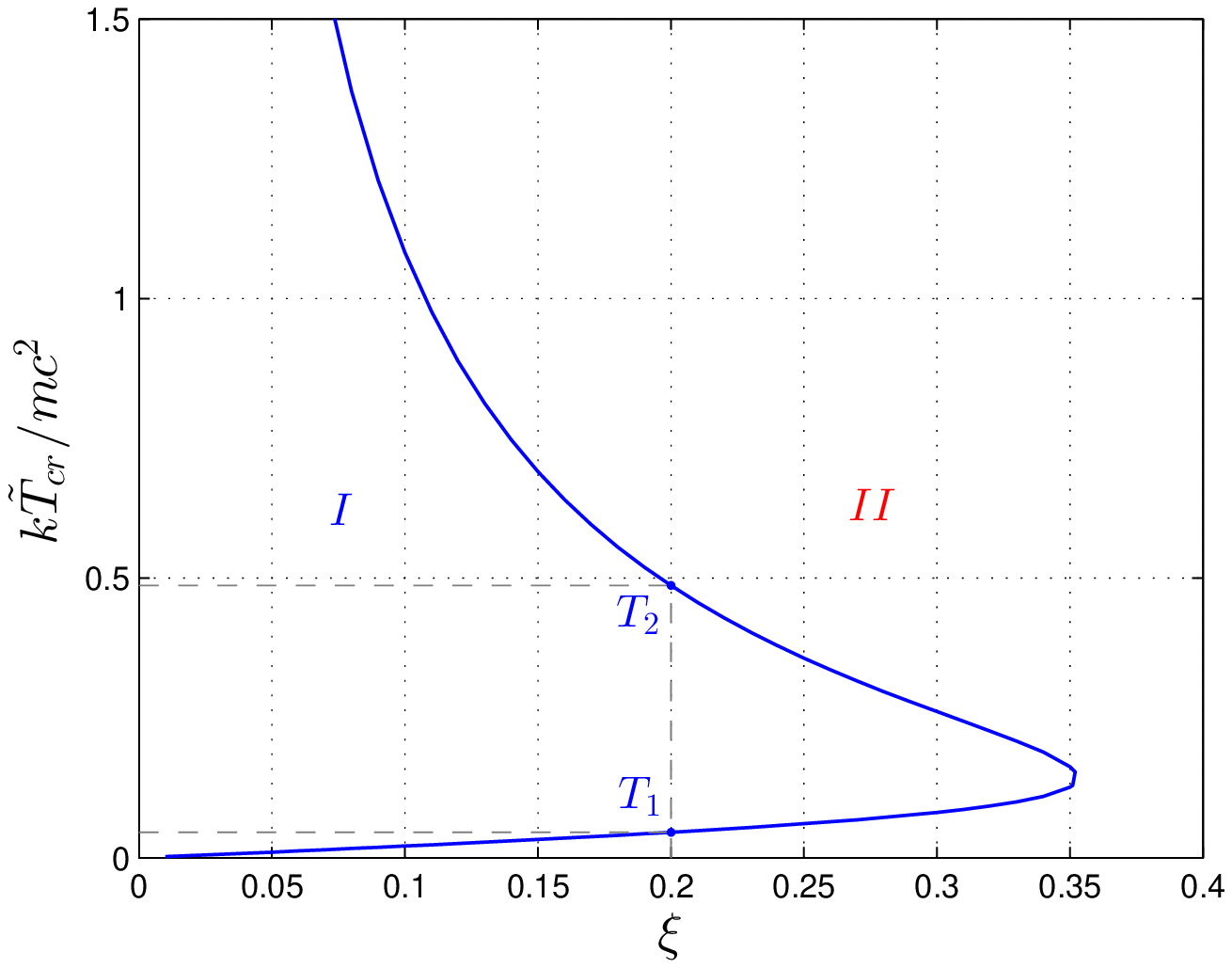}  } 
	\subfigure[]{ \label{fig:Tcr_chemical}\includegraphics[scale = 0.4]{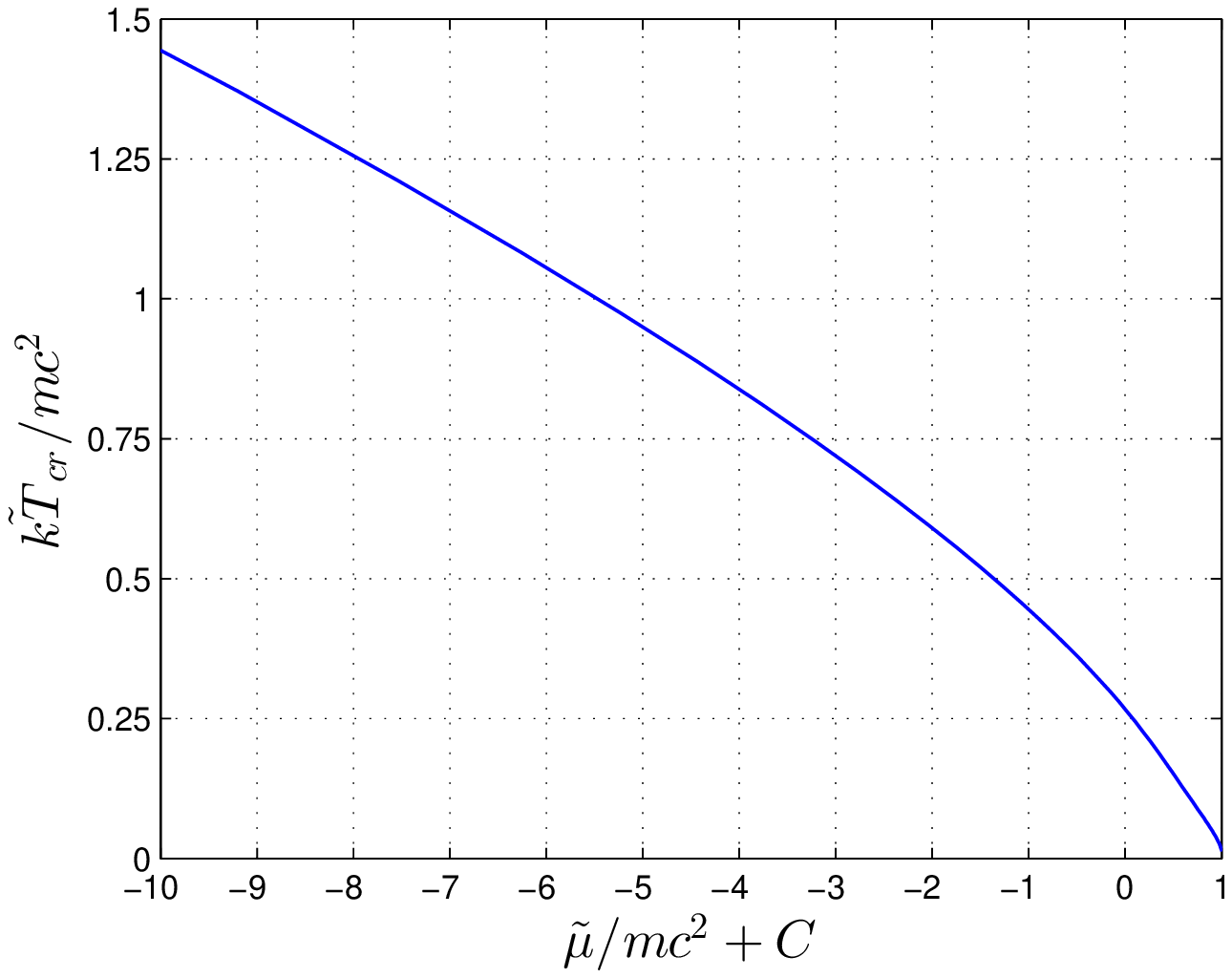} } 
	\caption{In (a) is plotted the critical Tolman temperature with respect to $\xi$. Equilibria exist only inside region $I$. It is evident that for every value of total rest mass and radius there are two critical temperatures: $T_1(\xi)$ for the low energy gravothermal instability and $T_2(\xi)$ for the high energy gravothermal instability. At these values an instability in the \textit{canonical ensemble} sets in. Equilibria exist only for $T_1(\xi)<T<T_2(\xi)$. In (b) is plotted the critical Tolman temperature with respect to the chemical potential. The checmical potential increases with temperature decrease and may also acquire positive values, due to the particle rest mass, as explained in section \ref{sec:EOS}. The constant $C$ is equal to $C = \ln(8\pi m^4 c^3/\rho_0 h^3)$.
	\label{fig:Tcr}}
\end{center} 
\end{figure}

\begin{figure}[ht]
\begin{center}
	\subfigure[]{ \label{fig:Rcr_can}\includegraphics[scale = 0.4]{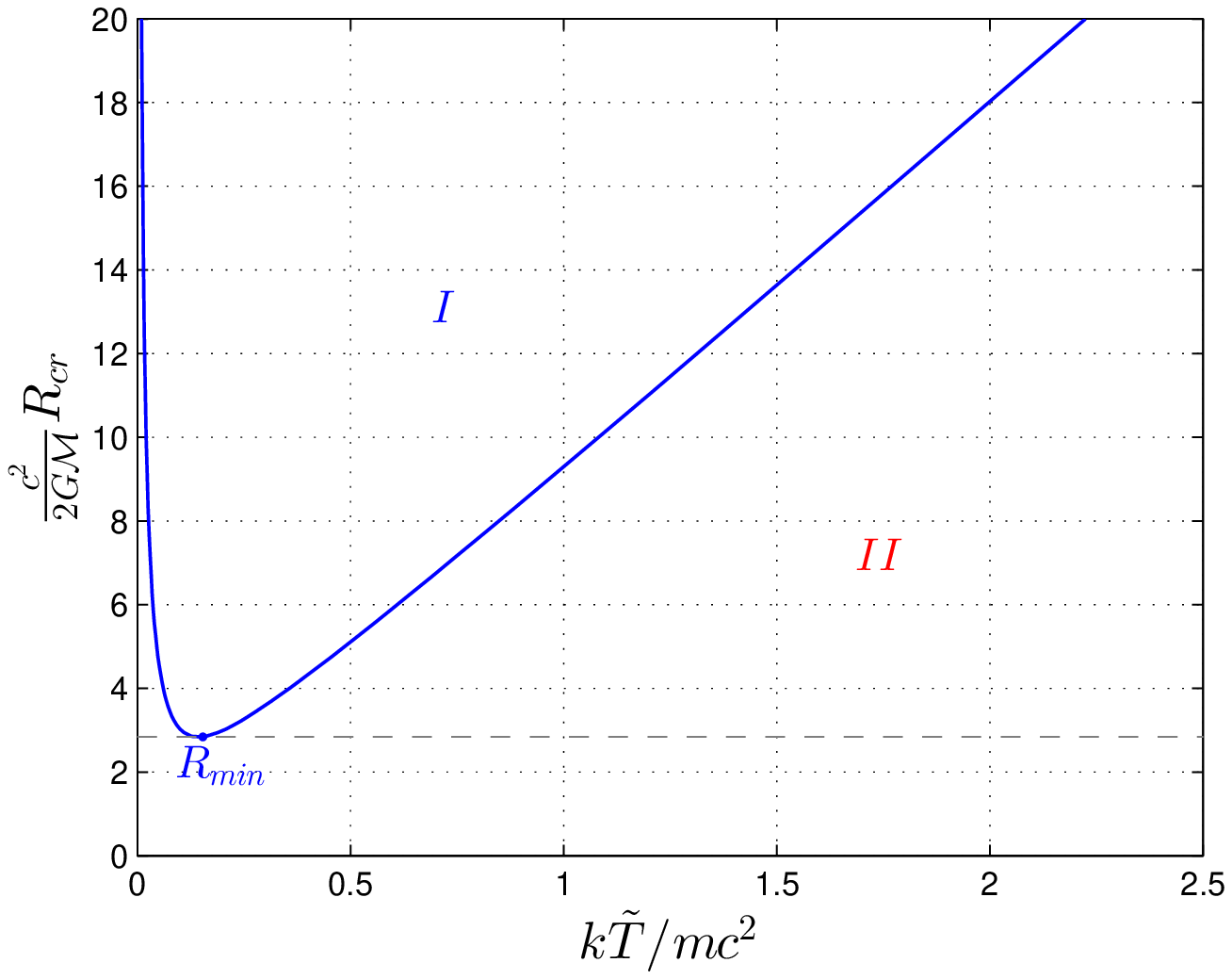}  } 
	\subfigure[]{ \label{fig:DC_can}\includegraphics[scale = 0.4]{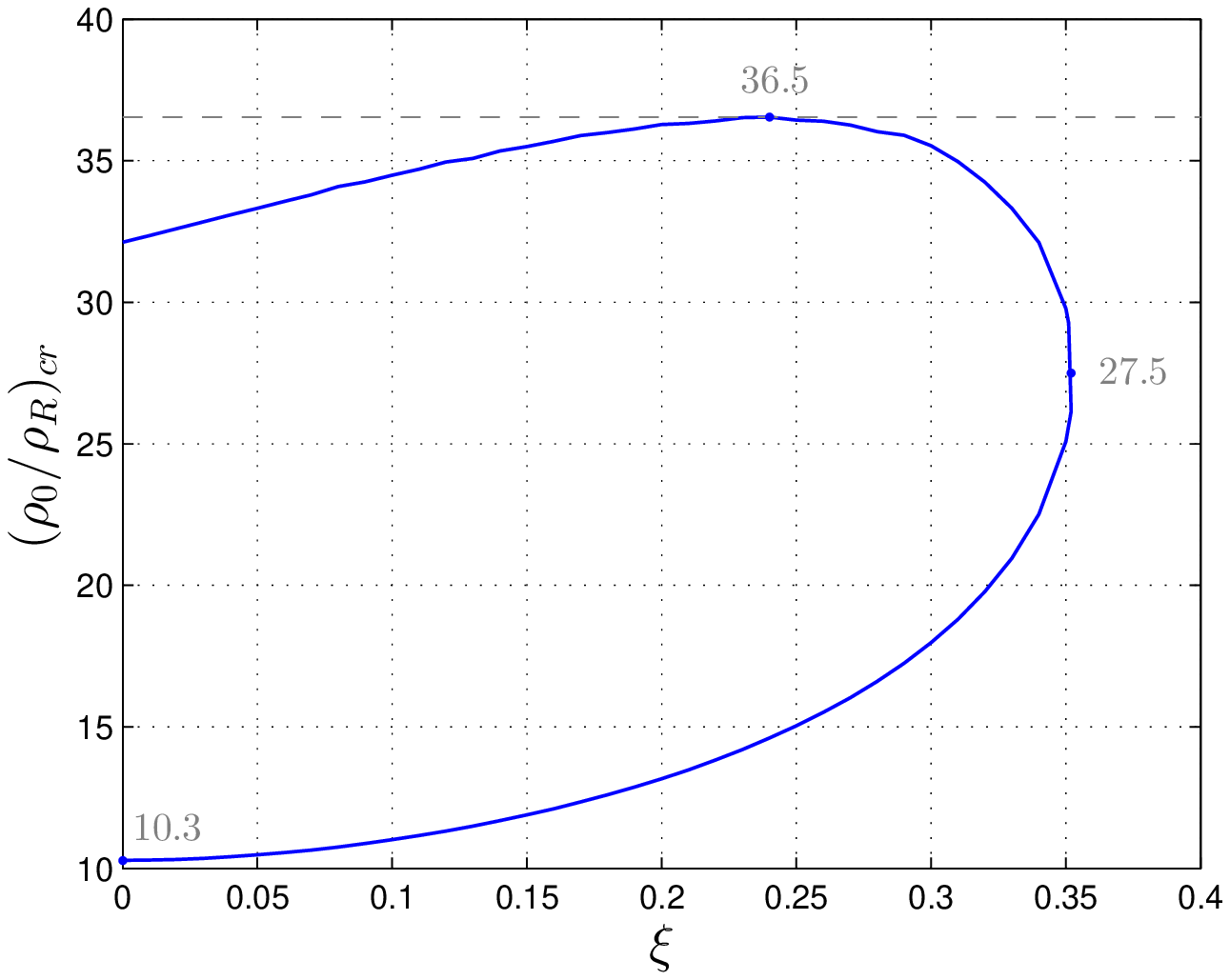} } 
	\caption{The critical radius and critical density contrast in the \textit{canonical ensemble}. The ultimate minimum of critical radius corresponds to $\xi_{max}$ and equals $R_{min} = 2.37R_S$. In (a), equilibria exist only inside region $I$. In (b), values above $27.5$ correspond to the low energy instability and values below to the high energy instability. An instability in the canonical ensemble sets in at each value of $R_{cr}$ and $(\rho_0/\rho_R)_{cr}$.
	\label{fig:R_DC_can}}
\end{center} 
\end{figure}

In Figure \ref{fig:Tcr} is plotted the critical Tolman temperature with respect to $\xi$. It is evident that for every value of total rest mass and radius there are two critical temperatures. Equilibria exist only in between these critical values. The upper value corresponds to the high energy gravothermal instability, while the lower value to the low energy gravothermal instability, both in the canonical ensemble.

In Figure \ref{fig:Rcr_can} is plotted the critical radius in the canonical ensemble with respect to temperature for fixed total rest mass. It presents an ultimate minimum
\begin{equation}
		R_{min} = 2.37R_S\; \mbox{, (canonical ensemble)}
\end{equation}
at $\xi = \xi_{max}$, where $k\tilde{T} = 0.19mc^2$.

In Figure \ref{fig:DC_can} is plotted the critical density contrast in the canonical ensemble with respect to $\xi$. At each of these values a gravothermal instability in the canonical ensemble sets in. The critical density contrast of the low energy instability varies in the interval $[27.5,36.5]$, while the high energy instability in the interval $[10.3,27.5]$. The ultimate maximum value $(\rho_0/\rho_R)_{max} = 36.5$ corresponds to $\xi = 0.24$ and $k\tilde{T} = 0.058mc^2$.

\section{Neutron Stars}\label{sec:NS}
	
Neutron stars are compact objects so dense that General Relativity comes into play. 
They are the remnants of core-collapse supernovae \cite{Baade:1934,Bethe:1985}. The initial core, called protoneutron star, is very hot $T\sim (50-100)MeV$ and is believed to cool rapidly by various processes of neutrino emission \cite{RevModPhys.64.1133,Page:2004fy,Yakovlev:2004iq,Page:2009} to temperatures $T \sim (0.1-100)keV$ depending on the core model and the stage of neutron core.

Neutron stars are composed \cite{Haensel:2007} of a dense, thick core, a thin crust and outer very thin envelope and atmosphere. The core determines the upper mass limit and the size of the star. It may be subdivided in the inner and outer cores. The inner core is ultra-dense with $\rho \geq 2\rho_N$ where $\rho_N = 2.8\cdot 10^{14} gr/cm^3$ is the normal nuclear density. Since these densities are unreachable from present laboratory experiments, its exact temperature, state of matter and therefore equation of state remains at present a mystery. Models vary \cite{Lattimer:2000nx,Haensel:2007} from (superfluid) $npe\mu$ gas, hyperons, Bose-Einstein condensates, kaons and pions to strange matter, deconfined quarks and Quark-Gluon-Plasma. The outer core is consisted mainly of neutrons although some protons, electrons and muons are present that prevent neutron decay. The crust consists of heavy nuclei and near the matching region with the outer core free neutrons are also present. The matching region is situated at density about \cite{Haensel:2007}
\begin{equation}\label{eq:rho_R_v}
	\rho_R \simeq \frac{\rho_N}{2} = 1.4\cdot 10^{14}\frac{gr}{cm^3},
\end{equation} 
where heavy nuclei can no longer exist. At this density a phase transition occurs towards the $npe\mu$ gas, through the capture of electrons by protons. 
	
Oppenheimer and Volkoff \cite{Oppenheimer:1939} calculated the upper mass limit of neutron cores assuming very cold, completely degenerate ideal gas, corresponding to the limiting case (\ref{eq:limit_Q}) of section \ref{sec:EOS}. We worked on the completely opposite limiting case (\ref{eq:limit_C}). Their result was $M_{OV} = 0.71M_\odot$. Including also protons, electrons and muons in $\beta$-equilibrium does not affect this upper mass limit \cite{Harrison:1958}. This value turned out to be very low compared to observations and was later, after the discovery of the first neutron star \cite{Hewish:1968}, regarded as a proof of the fact that nuclear forces have repulsive effects at supra-nuclear densities \cite{Cameron:1959,Zeldovich:1962}. When nuclear forces are taken into account the limit increases and may reach two solar masses depending on the model \cite{Haensel:2007,Lattimer:2012}. 	
	
Observations of neutron stars indicate that they have masses of about $1.4M_\odot$ and radius $10-15km$. The highest accurately measured masses are about $M = 2M_\odot$ observed very recently \cite{Antoniadis:2013pzd,Demorest:2010}. The existence of ultra-heavy neutron stars of about or more than two solar masses is of extreme importance for Physics. The upper mass limit depends on the equation of state inside the core which, as we already remarked, is unknown. Ultra-heavy neutron stars will rule out several models of ultra-dense matter. In addition, ultra-heavy neutron stars lie on the regime where General Relativity may become invalid and quantum corrections or modifications of the theory may be necessary. 

One may wonder, what is the upper mass limit of neutron cores if only thermal pressure and not degeneracy pressure is taken into account, so that the classical ideal gas may be applied. Although this classical limit may not be valid for neutron stars since they are believed to be cold, it is a legitimate theoretical question whose answer will shed light on the ability of thermal energy to halt gravitational collapse. In addition, the fact that at the beginning of their lives neutron stars are hot, together with the high uncertainty regarding the conditions and composition of the inner core, justify this investigation.

\begin{figure}[ht]
\begin{center}
	\includegraphics[scale = 0.55]{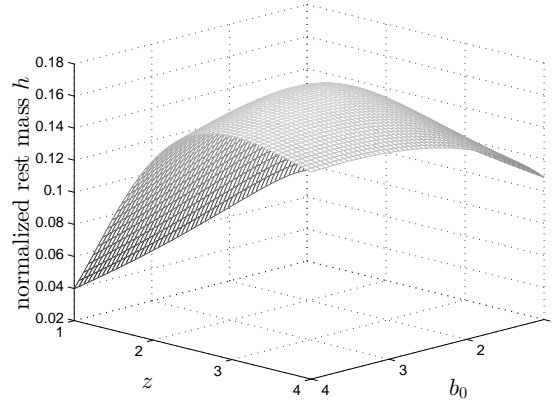}  	
	\caption{The normalized rest mass $h = \mathcal{M} \sqrt{\rho_R(16\pi G^3/c^6)}$ with respect to the central inverse temperature $b_0 = mc^2/kT_0$ and the quantity $z = R\sqrt{4\pi G \rho_0b_0/c^2}$. It presents a global maximum at $h_{max} = 0.165$.
	\label{fig:Mrest_neutron_3D}}
\end{center} 
\end{figure}

So, let apply the system (\ref{eq:TOV_ND}-\ref{eq:mass_ND}) at the core of the star, which is surrounded by the crust at $r=R$. Therefore, we have to solve the system (\ref{eq:TOV_ND}-\ref{eq:mass_ND}) assuming a constant mass density $\rho_R$ at the edge, given by (\ref{eq:rho_R_v}).
We have to generate the $M-R$ curve in order to study the stability of the sphere. However, there are many different $M-R$ curves, each one for every fixed value of total rest mass $\mathcal{M}$.
In order to calculate the upper limit of all these $M-R$ curves, we define the normalized total rest mass:
\begin{equation}
	h = 2 u_r b_0^{\frac{3}{2}}e^{-\frac{y}{2}} \equiv \mathcal{M} \sqrt{\frac{16\pi G^3}{c^6}\rho_R},
\end{equation} 
where $u_r = \mathcal{M}G\frac{b_0}{c^2}\sqrt{4\pi G \rho_0\frac{b_0}{c^2}}$.

In Figure \ref{fig:Mrest_neutron_3D} is plotted $h$ with respect to $(b_0,z)$. It presents a global maximum at $h_{max} = 0.165$ that gives:
\begin{equation}
	\mathcal{M}_{max} = 1.55 M_\odot,
\end{equation}
where $M_\odot$ is the solar mass. 
In order to calculate $M_{max}$ the system is solved numerically choosing pairs $(z,b_0)$ such, that $h$ is kept fixed. This is achieved by developing an appropriate computer program. The procedure is repeated for various values of $h$. Assuming the fixed value for $\rho_R$, this procedure accounts for solving for various rest masses. The system may slightly move across various $M-R$ curves, because of the random particle exchange between the core and the crust, however the total mass cannot exceed the upper limit of all mass maxima for every $h$.

\begin{figure}[ht]
\begin{center}
	\subfigure[]{ \label{fig:Mplus_R}\includegraphics[scale = 0.4]{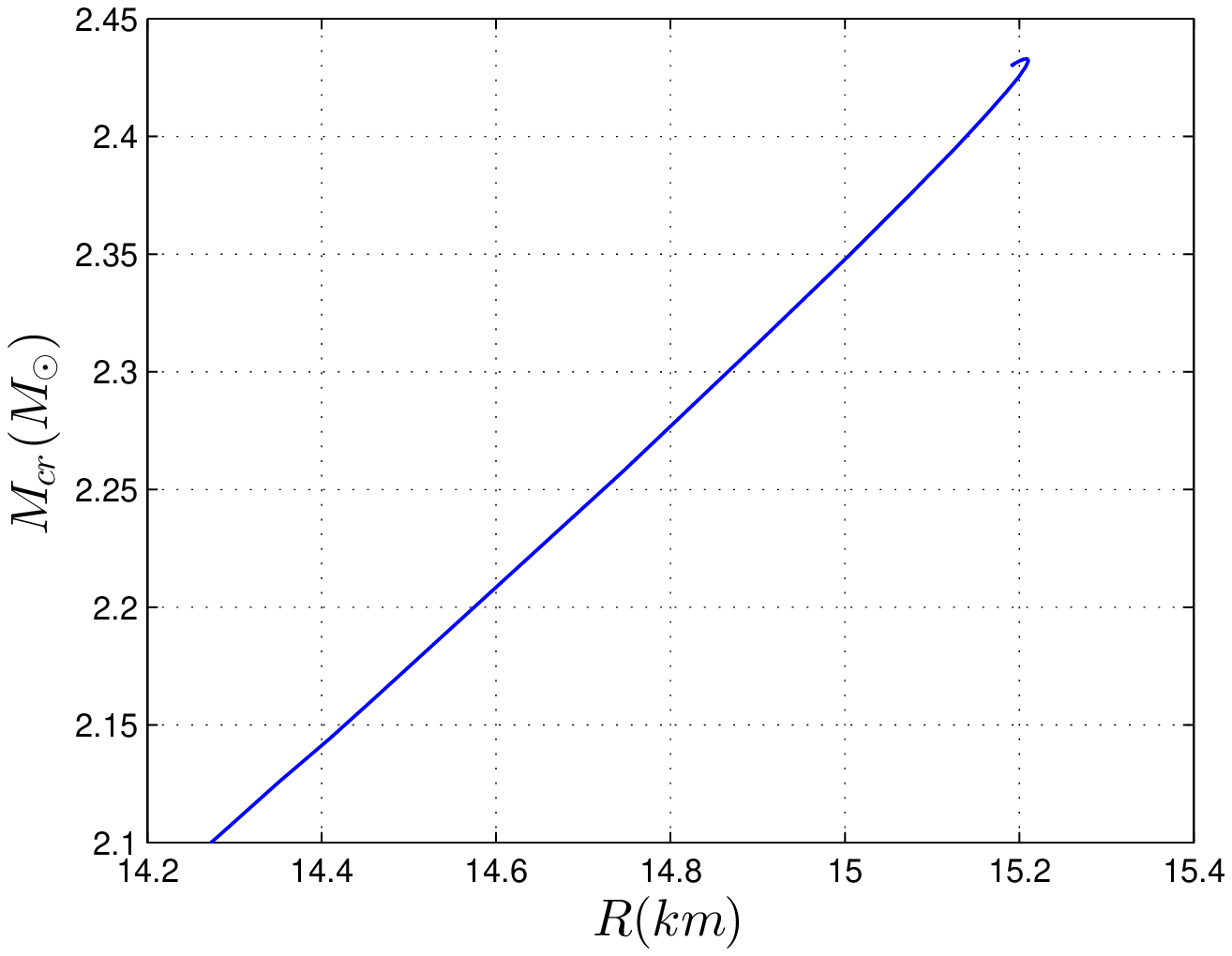} } 
	\subfigure[]{ \label{fig:Mplus_ksi}\includegraphics[scale = 0.4]{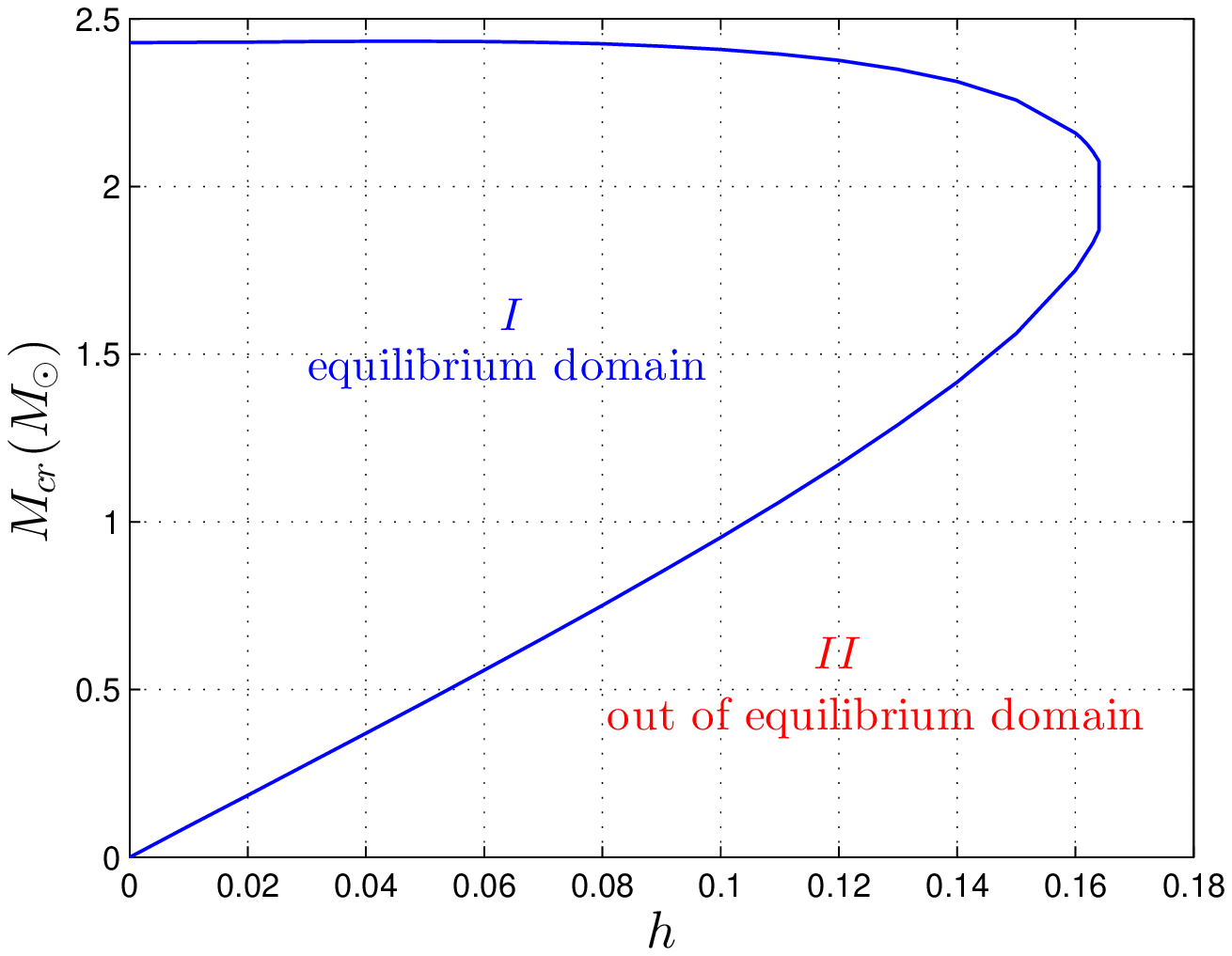} } 
	\caption{The critical maximum total masses $M_{cr}$ of hot neutron cores with respect to the radius at which they appear in Figure (a) and to the rest mass $h = \mathcal{M} \sqrt{\rho_R(16\pi G^3/c^6)}$ in (b). In (b), equilibria may be found only inside region $I$. The edge denisty is fixed to the value $\rho_R = 1.4\cdot 10^{14}gr/cm^3$. It is evident that critical masses present an upper limit $M_{max} = 2.43M_\odot$ at radius $R = 15.2km$.
	\label{fig:Mplus_1}}
\end{center} 
\end{figure}
	
\begin{figure}[ht]
\begin{center}
	\includegraphics[scale = 0.55]{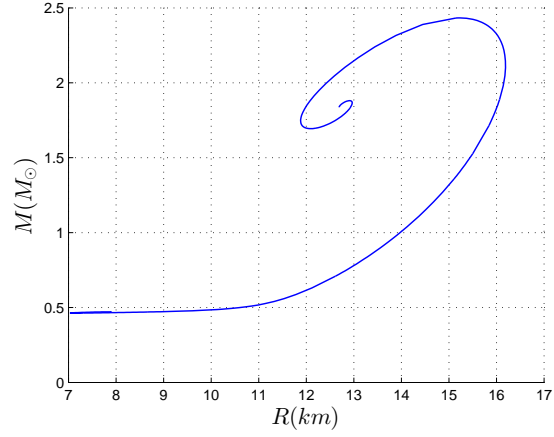}  	
	\caption{The mass-radius diagram for the marginal equilibrium of hot neutron cores with fixed edge density $\rho_R = 1.4\cdot 10^{14}gr/cm^3$ with total mass equal to the upper mass limit $M_{max} = 2.43M_\odot$.
	\label{fig:Spiral_neutron}}
\end{center} 
\end{figure}

\begin{figure}[ht]
\begin{center}
	\subfigure[]{ \label{fig:Mplus_T}\includegraphics[scale = 0.4]{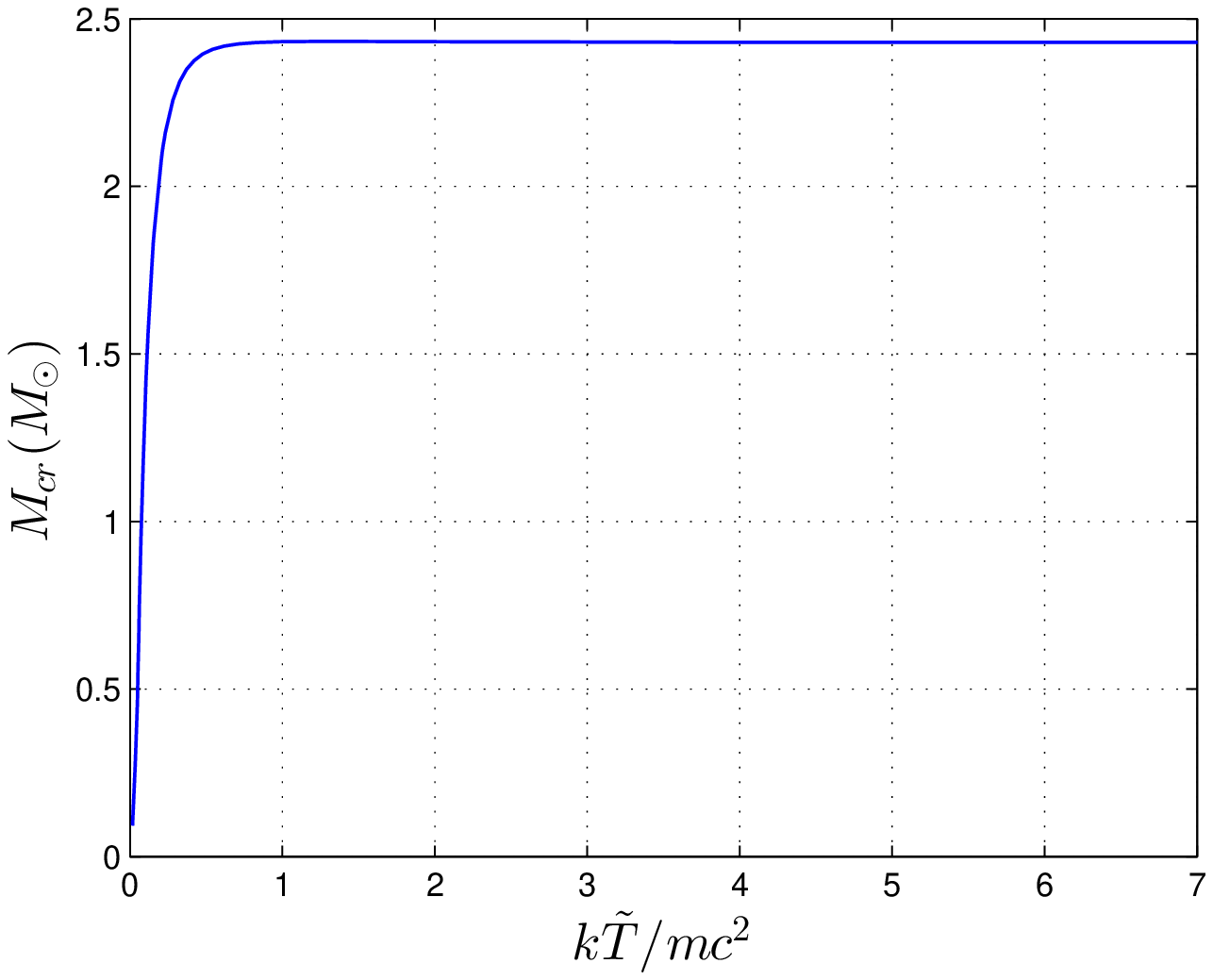} } 
	\subfigure[]{ \label{fig:Mplus_mu}\includegraphics[scale = 0.4]{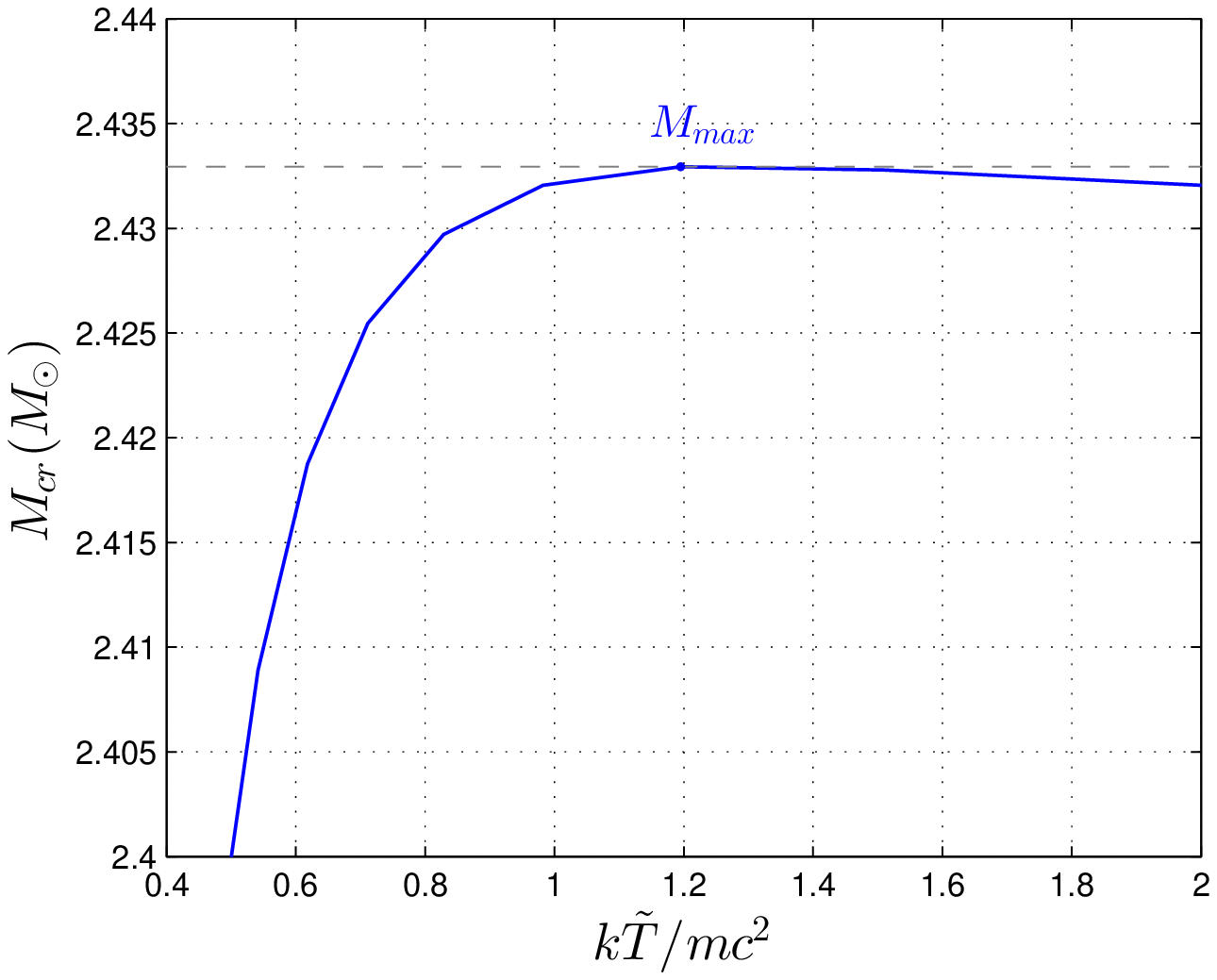} } 
	\caption{The critical maximum total masses $M_{cr}$ of hot neutron cores with respect to the corresponding Tolman temperature. The edge denisty is fixed to the value $\rho_R = 1.4\cdot 10^{14}gr/cm^3$. It is evident that critical masses present an upper limit $M_{max} = 2.43M_\odot$ with temperature $k\tilde{T}/mc^2 = 1.2$.
	\label{fig:Mplus_T_NS}}
\end{center} 
\end{figure}

In Figure \ref{fig:Mplus_R} are plotted the total mass maxima, denoted $M_{cr}$, with respect to the radius at which they appear. It is evident an upper limit of all these maxima. Its value is
\begin{equation}\label{eq:M_max_NS}
	M_{max} = 2.43M_\odot
\end{equation}
at radius
\begin{equation}
	R|_{M_{max}} = 15.2km.
\end{equation}
The $M-R$ curve of this equilibrium is plotted in Figure \ref{fig:Spiral_neutron}.

These values for $M_{max}$ and $R$ are unexpectedly greatly matching observations! However, the temperature corresponding to the limiting mass is very high: 
\begin{equation}
\left.	\frac{k\tilde{T}}{mc^2}\right|_{M_{max}} = 1.2
\end{equation}
The critical masses with respect to Tolman temperature is plotted in Figure \ref{fig:Mplus_T_NS}.
Incidentally, the maximum mass corresponding to $\mathcal{M}_{max}$, is 
\begin{equation}
	M_{max}|_{\mathcal{M}_{max}} = 1.97M_\odot
\end{equation}
a value almost equal to the upper mass imposed by observations, namely $M_{obs} \leq 2M_\odot$. However, even to this value corresponds a high temperature of:
\begin{equation}
\left.	\frac{k\tilde{T}}{mc^2}\right|_{\mathcal{M}_{max}} = 0.18.
\end{equation}
The corresponding radius is
\begin{equation}
	R|_{\mathcal{M}_{max}} = 13.6km.
\end{equation}
The chemical potential for each case is:
\begin{eqnarray*}
\left.	\frac{\tilde{\mu}}{mc^2}\right|_{M_{max}} = -9.8 \\
\left.	\frac{\tilde{\mu}}{mc^2}\right|_{\mathcal{M}_{max}} = -1.7
\end{eqnarray*}

Comparing the maximum value obtained here $M_{max} = 2.4M_\odot$ with the Oppenheimer-Volkoff value\footnote{One might object that the Oppenheimer-Volokoff calculation involves a different boundary condition, namely vanishing edge pressure. However, as I will show in a subsequent work, the boundary condition assumed here $\rho_R =\rho_N/2$, only slightly changes the OV result giving a maximum mass of $0.68 M_\odot$.} $M_{OV} = 0.7M_\odot$, it is evident that thermal pressure is much more effective in holding gravitational collapse than degeneracy pressure. The corresponding temperature for the thermal case is, however, very high. On the other hand the mass and radius values predicted are so accurate, with respect to observation, that one should normally wonder if this is just a coincidence. After all, neutron stars are believed to be hot at the beginning of their life \cite{Burrows:1986me,Prakash:1997}. In addition, the composition and conditions inside the inner core are unknown and also the densities of ultra-heavy neutron stars, heavier than two solar masses, are so high, that unknown processes coming from (quantum) extensions of General Relativity may intervene \cite{PhysRevLett.70.2220,PhysRevD.54.1474,PhysRevLett.105.151102,Landulfo:2014wra}. Overall, this result may apply to protoneutron stars and underlines the necessity to carry out an analysis of the full problem, with both degeneracy and thermal pressure taken into account and thus to calculate the mass limit through the whole range of temperature. I perform this analysis in a separate work \cite{Roupas:2014hqa}.

\section{Conclusions}\label{sec:Con}

Consider an ideal gas, bounded inside a spherical shell	with fixed radius. Imagine you begin to heat it up. Particles will move faster enhancing the tendency of the gas to expand and increasing the pressure to the walls. However, since the energy, even heat, has mass, at some point General Relativity will begin to take over. The absorbed heat, turned into internal kinetic energy, will cause gravitational collapse of the gas! This is the essence of the \textit{high energy relativistic gravothermal instability} presented here.

On the other hand, if you absorb energy from the gas, at some point it will also collapse under self-gravity. But, this time the reason will be the lack of thermal energy and not the excess of it. This is the \textit{low energy relativistic gravothermal instability} presented here. It is in fact the relativistic generalization of Antonov instability.

These relativistic gravothermal instabilities make themselves evident in the temperature versus energy series of equilibria as a double spiral. This was demonstrated in Figures \ref{fig:DoubleSpiral_1} and \ref{fig:DoubleSpiral_2}.
The turning points, i.e. the points at which these instabilities set in, are found to depend on the total rest mass over radius ratio $\xi = 2G\mathcal{M}/Rc^2$. An ultimate upper limit of $\xi$ is reported, namely $\xi_{max} = 0.35$. At this value both turning points along with all equilibria merge into a single equilibrium state. 

Thermal equilibrium imposes Tolman relation for the temperature and chemical potential. This Tolman-Ehrenfest effect is taken into account for all calculations and is shown in Figures \ref{fig:Tolman_Ehrenfest} and \ref{fig:Chemical_Potential} for two equilibria.

General relativistic effects render the system more unstable. As the relativistic control parameter $\xi$ is increased the double spiral shrinks (maximum energy and temperature decrease, while minimum energy and temperature increase) and completely disappears for $\xi > \xi_{max}$.

An ultimate upper limit of all mass maxima (one maximum for every one $\xi$) is also reported: $M_{max} = 0.49M_S$. It corresponds to $\xi = 0.08$ and Tolman temperature $k\tilde{T}/mc^2 = 1.3$. The critical masses curve for both instabilities types were given in Figures \ref{fig:Ecr} and \ref{fig:Mcr_micro}.

I stress that both gravothermal instabilities types, in the microcanonical ensemble set in when the specific heat becomes zero, with the equilibria lying on the negative specific heat side being stable, while the ones at the positive side being unstable! This indicates that the collapse in both types follows similar pattern which leads to a core-halo structure as indicated by Lynden-Bell \& Wood \cite{Bell:1968} for Antonov instability, which presents similar behaviour of specific heat.

The available radii are also constrained by the relativistic instabilities. In the microcanonical ensemble and for negative gravothermal energy the radius of isothermal spheres is constrained by two limits $R_2(\xi)<R<R_1(\xi)$ with $R_1(\xi)$ corresponding to the low energy instability and $R_2(\xi)$ to the high energy instability. The ultimate minimum radius in the microcanonical ensemble is $R_{min} = 2.03R_S$ corresponding to $\xi = 0.08$. The critical radii in the microcanonical ensemble were plotted in Figure \ref{fig:Rcr_micro}. In the canonical ensemble there is only a minimum radius with an ultimate value $R_{min} = 2.37R_S$ as demonstrated in Figure \ref{fig:Rcr_can}. 

The critical density contrast is also calculated in both ensembles and given in Figures \ref{fig:DC_micro} and
\ref{fig:DC_can}.

Finally, the maximum mass limit of non-degenerate neutron cores is calculated. The result $M_{max} = 2.4M_\odot$
 with $R = 15.2km$ at high temperature suggests that thermal pressure alone is more 
 effective in halting gravitational collapse than degeneracy pressure alone for which the limit is 
 $M_{OV} = 0.7M_\odot$ at zero temperature. Although neutron cores of actual neutron stars 
 are believed to be completely degenerate and cold, neutron stars at the beginning of their life, called `protoneutron stars', are believed to be very hot. This maximum mass and corresponding radius, calculated here, match so nicely 
  observations that one might naturally wonder if it is just a coincidence. The dependence of mass limit to 
  temperature is given in Figure \ref{fig:Mplus_T_NS}. However, at low temperatures the classical limit is invalid 
  and degeneracy pressure should be taken into account. The complete analysis taking into account both thermal and 
  degeneracy pressure is carried out in a separate work \cite{Roupas:2014hqa}.

\section*{Acknowledgements}

\indent This work was mostly elaborated during my stay at Charles University of Prague. I would like to thank the Relativity Group for their hospitality. 

\appendix
\section{}\label{app:A}

It is shown here, that, in the Newtonian limit, the following system of equations (namely the TOV, mass, Tolman equations and the relativistic classical ideal gas equation of state):
\begin{eqnarray}
\label{app:TOV}
\frac{d P}{dr} &= - \left({\frac{P}{c^2}} + \rho\right) {\left(\frac{G\hat{M}}{r^2} + 4\pi G \frac{P}{c^2} r\right) \left(1 - \frac{2G\hat{M}}{rc^2} \right)^{-1} } 
\\
\label{app:mass}
\frac{d \hat{M}}{dr} &= 4\pi \rho r^2,	
\\
\label{app:tolman}
	\frac{db}{dr} &=  	-\frac{b}{P+\rho c^2} \frac{dP}{dr}, 
\\
\label{app:eos}
	P(r) &= \frac{1}{b(r)(1+\mathcal{F}(b(r)))}\rho(r) c^2,
\end{eqnarray}
reduces to the well-known Emden equation:
\begin{equation}\label{app:Emden}
	\frac{1}{r^2}\frac{d}{dr}\left( r^2\frac{d \phi(r)}{dr}\right) = 4\pi G \rho_0 e^{-\frac{m}{k T_N} (\phi(r)-\phi(0))},
\end{equation}
where $\phi(r)$ is the Newtonian potential and $T_N$ is the Newtonian temperature.
Apparently, the Emden equation is just the Poisson equation for a classical ideal gas in spherical coordinates. One may wonder where is the gravitational potential hidden in the relativistic framework (\ref{app:TOV}-\ref{app:eos}). As we will see it emerges out of the relativistic temperature $T(r)$ in the weak-field limit!

The time-time component of the metric $g_{tt}$ is written in the weak-field limit as:
\begin{equation}
	g_{tt}\simeq 1+ \frac{2\phi}{c^2},
\end{equation}
Let also be
\begin{equation}
	g_{tt} = e^{\nu(r)}
\end{equation}
as usual in the spherically symmetric case. Then, in the weak-field limit, it is:
\begin{equation}\label{app:nu}
	\nu \simeq \frac{2\phi}{c^2}.
\end{equation}
As I have already noted in section \ref{sec:TE}, equation (\ref{app:tolman}) is the differentiated form of Tolman's relation:
\begin{equation}\label{app:tolman_rel}
	T(r)e^{\frac{\nu}{2}} = \tilde{T} \equiv const,
\end{equation}
where, in the present text, we have be calling the constant $\tilde{T}$ as the Tolman temperature. In the weak-field limit it is therefore through equation (\ref{app:nu}):
\begin{equation}\label{app:temp}
	\frac{1}{kT(r)} \simeq \frac{1}{k\tilde{T}}\left( 1 + \frac{\phi(r)}{c^2}\right)
\end{equation}
Using also the definition $b(r) = mc^2/kT(r)$, we get:
\begin{equation}\label{app:b}
	b(r)-b(0)\simeq \frac{m}{k\tilde{T}}(\phi(r)-\phi(0))
\end{equation}
Now, since as we have already seen in section \ref{sec:EOS_cl}, the equation of state (\ref{app:eos}) reduces in the Newtonian limit to the equation:
\begin{equation}\label{app:eos_N}
	P \simeq \frac{\rho c^2}{b},
\end{equation}
the Tolman equation (\ref{app:tolman}) becomes:
\begin{equation}\label{app:bprime}
	\frac{b'}{b} \simeq -\frac{\frac{\rho'}{b} - \rho\frac{b'}{b^2}}{\rho\left(\frac{1}{b} + 1\right)}
	\Rightarrow b' \simeq -\frac{\rho'}{\rho} 
\end{equation}
and therefore
\begin{equation}\label{app:rho}
	\rho(r) \simeq \rho_0 e^{-(b(r)-b(0))}.
\end{equation}
Substituting equation (\ref{app:b}) we finally get:
\begin{equation}\label{app:rho_phi}
	\rho(r) \simeq \rho_0 e^{-\frac{m}{k\tilde{T}}(\phi(r)-\phi(0))}.
\end{equation}
It is evident that the Tolman temperature $\tilde{T}$ should be identified with the Newtonian temperature $T_N$.

Now, applying the non-(special)relativistic limit $c\rightarrow\infty$ to TOV equation (\ref{app:TOV}), in top of the weak-field limit, which gave equation (\ref{app:rho_phi}), it is straightforward to obtain Emden equation. For $c\rightarrow\infty$ we have from equation (\ref{app:temp})) that 
\begin{equation}\label{app:temp_c}
	\lim_{c\rightarrow\infty}T(r) = \tilde{T} = const.
\end{equation} 
and therefore the equation of state (\ref{app:eos_N}) gives:
\begin{equation}
	P \simeq \rho \frac{k\tilde{T}}{m} \Rightarrow P' = \rho' \frac{k\tilde{T}}{m}.
\end{equation}
Finally, TOV equation (\ref{app:TOV}) in the limit $c\rightarrow\infty$ becomes:
\begin{equation}\label{app:final}
	\frac{k\tilde{T}}{m}\frac{d\rho}{dr}  \simeq - \rho\frac{G\hat{M}}{r^2} \Rightarrow
	r^2\frac{d}{dr}\left( \frac{k\tilde{T}}{m}\ln (\rho)\right) = -G\hat{M}.
\end{equation}
We are finally led to the Emden equation (\ref{app:Emden}) by differentiating, substituting the mass equation (\ref{app:mass}), then substituting equation (\ref{app:rho_phi}) and identifying $\tilde{T} \equiv T_N$.

\section*{References}

\bibliography{Roupas_Relativistic_Gravothermal}

\begin{thebibliography}{10}

\bibitem{Antonov:1962}
V.~Antonov,
\newblock Vest. Leningrad Univ. {\bf 7}, 135 (1962).

\bibitem{Bell:1968}
D.~Lynden-Bell and R.~Wood,
\newblock MNRAS {\bf 138}, 495 (1968).

\bibitem{Padman:1989}
T.~Padmanabhan,
\newblock ApJS {\bf 71}, 651 (1989).

\bibitem{Axenides:2012bf}
M.~Axenides, G.~Georgiou, and Z.~Roupas,
\newblock Phys.Rev. {\bf D86}, 104005 (2012), arXiv:1206.2839.

\bibitem{Chavanis:2001ah}
P.-H. Chavanis,
\newblock A\&A {\bf 381}, 709 (2002), arXiv:astro-ph/0108230.

\bibitem{Tolman:1930}
R.~Tolman,
\newblock Phys. Rev. {\bf 35}, 904 (1930).

\bibitem{Padman:1990}
T.~Padmanabhan,
\newblock Phys. Rep. {\bf 188}, 285 (1990).

\bibitem{Roupas:2014nea}
Z.~Roupas,
\newblock Class.Quant.Grav. {\bf 32}, 119501 (2015), arXiv:1411.0325.

\bibitem{Roupas:2013nt}
Z.~Roupas,
\newblock Class.Quant.Grav. {\bf 30}, 115018 (2013), arXiv:1301.3686.

\bibitem{Tolman:1939}
R.~C. Tolman,
\newblock Phys. Rev. {\bf 55}, 364 (1939).

\bibitem{Oppenheimer:1939}
J.~R. Oppenheimer and G.~M. Volkoff,
\newblock Phys. Rev. {\bf 55}, 374 (1939).

\bibitem{Tolman-Ehrenfest:1930}
R.~C. Tolman and P.~Ehrenfest,
\newblock Phys. Rev. {\bf 36}, 1791 (1930).

\bibitem{Klein:1949}
O.~{Klein},
\newblock Reviews of Modern Physics {\bf 21}, 531 (1949).

\bibitem{Merafina:1989}
{{Merafina}, M. and {Ruffini}, R.},
\newblock A\&A {\bf 221}, 4 (1989).

\bibitem{Burrows:1986me}
A.~Burrows and J.~M. Lattimer,
\newblock Astrophys.J. {\bf 307}, 178 (1986).

\bibitem{Prakash:1997}
M.~Prakash {\em et~al.},
\newblock Physics Reports {\bf 280}, 1 (1997).

\bibitem{Chandra:1938}
S.~Chandrasekhar,
\newblock {\em An introduction to the study of stellar structure} (Chicago,
  iLLINOIS, 1938).

\bibitem{Chavanis:2001hd}
P.-H. Chavanis,
\newblock A\& A {\bf 381}, 340 (2001), arXiv:astro-ph/0103159.

\bibitem{Green:2013ica}
S.~R. Green, J.~S. Schiffrin, and R.~M. Wald,
\newblock Class.Quant.Grav. {\bf 31}, 035023 (2014), arXiv:1309.0177.

\bibitem{Weinberg}
S.~Weinberg,
\newblock {\em Gravitation and Cosmology: Principles and applications of the
  General Theory of Relativity} (Wiley New York, 1972).

\bibitem{Israel:1963}
W.~Israel,
\newblock J. Math. Phys. {\bf 4}, 1163 (1963).

\bibitem{Cocke:1965}
W.~J. {Cocke},
\newblock Ann. Inst. Henri Poincar\'e {\bf 2}, 283 (1965).

\bibitem{SWZ:1981}
R.~Sorkin, R.~Wald, and Z.~Jiu,
\newblock General Relativity and Gravitation {\bf 13}, 1127 (1981).

\bibitem{Gao:2011hh}
S.~Gao,
\newblock Phys.Rev. {\bf D84}, 104023 (2011), arXiv:1109.2804.

\bibitem{Gao:2012add}
S.~Gao,
\newblock Phys. Rev. D {\bf 85}, 027503 (2012).

\bibitem{Fang:2013oka}
X.~Fang and S.~Gao,
\newblock Phys.Rev. {\bf D90}, 044013 (2014), arXiv:1311.6899.

\bibitem{Roupas:2014hqa}
Z.~Roupas,
\newblock Phys.Rev. {\bf D91}, 023001 (2015), arxiv:1411.5203.

\bibitem{LyndenBell:1966bi}
D.~Lynden-Bell,
\newblock Mon.Not.Roy.Astron.Soc. {\bf 136}, 101 (1967).

\bibitem{King:1966fn}
I.~R. King,
\newblock Astron.J. {\bf 71}, 64 (1966).

\bibitem{Hjorth:1991}
J.~{Hjorth} and J.~{Madsen},
\newblock MNRAS {\bf 253}, 703 (1991).

\bibitem{Hjorth:1993}
J.~{Hjorth} and J.~{Madsen},
\newblock MNRAS {\bf 265}, 237 (1993).

\bibitem{Chavanis21051998}
P.-H. Chavanis and J.~Sommeria,
\newblock MNRAS {\bf 296}, 569 (1998).

\bibitem{Chavanis:2002gg}
P.-H. Chavanis,
\newblock Astron.Astrophys. {\bf 401}, 15 (2003), arXiv:astro-ph/0207080.

\bibitem{BisnovatyiKogan:2006cw}
G.~S. Bisnovatyi-Kogan and M.~Merafina,
\newblock Astrophys.J. {\bf 653}, 1445 (2006), arXiv:astro-ph/0603398.

\bibitem{Casetti:2012}
L.~{Casetti} and C.~{Nardini},
\newblock Phys. Rev. E {\bf 85}, 061105 (2012), arXiv:1203.1284.

\bibitem{Hawking:1976}
S.~W. Hawking,
\newblock Phys. Rev. D {\bf 13}, 191 (1976).

\bibitem{Gross:1982}
D.~J. Gross, M.~J. Perry, and L.~G. Yaffe,
\newblock Phys. Rev. D {\bf 25}, 330 (1982).

\bibitem{Pavon:1988}
D.~Pav\'on and P.~T. Landsberg,
\newblock General Relativity and Gravitation {\bf 20}, 457 (1988).

\bibitem{Haensel:2007}
A.~P. P.~Haensel and D.~Yakovlev,
\newblock {\em Neutron Stars I} (Springer, 2007).

\bibitem{Poincare}
H.~Poincar{\'e},
\newblock Acta. Math. {\bf 7}, 259 (1885).

\bibitem{Katz:1978}
T.~Katz,
\newblock MNRAS {\bf 183}, 765 (1978).

\bibitem{Axenides:2013npb}
M.~Axenides, G.~Georgiou, and Z.~Roupas,
\newblock Nucl. Phys. B {\bf 871}, 21 (2013), arXiv:1302.1977.

\bibitem{Axenides:2013hba}
M.~Axenides, G.~Georgiou, and Z.~Roupas,
\newblock J.Phys.Conf.Ser. {\bf 410}, 012130 (2013), arxiv:1303.4543.

\bibitem{Baade:1934}
W.~Baade and F.~Zwicky,
\newblock Phys. Rev. {\bf 45}, 138 (1934).

\bibitem{Bethe:1985}
H.~A. Bethe and J.~R. Wilson,
\newblock ApJ {\bf 295}, 14 (1985).

\bibitem{RevModPhys.64.1133}
C.~J. Pethick,
\newblock Rev. Mod. Phys. {\bf 64}, 1133 (1992).

\bibitem{Page:2004fy}
D.~Page, J.~M. Lattimer, M.~Prakash, and A.~W. Steiner,
\newblock Astrophys.J.Suppl. {\bf 155}, 623 (2004), arXiv:astro-ph/0403657.

\bibitem{Yakovlev:2004iq}
D.~G. Yakovlev and C.~Pethick,
\newblock Ann.Rev.Astron.Astrophys. {\bf 42}, 169 (2004),
  arXiv:astro-ph/0402143.

\bibitem{Page:2009}
D.~Page, J.~M. Lattimer, M.~Prakash, and A.~W. Steiner,
\newblock ApJ {\bf 707}, 1131 (2009), arXiv:0906.1621.

\bibitem{Lattimer:2000nx}
J.~Lattimer and M.~Prakash,
\newblock Astrophys.J. {\bf 550}, 426 (2001), arXiv:astro-ph/0002232.

\bibitem{Harrison:1958}
B.~Harrison, M.~Wakano, and J.~Wheeler,
\newblock La structure et \'evolution de l'univers , 124 (1958).

\bibitem{Hewish:1968}
A.~Hewish, S.~Bell, J.~Pilkington, P.~Scott, and R.~Collins,
\newblock Nature {\bf 709} (1968).

\bibitem{Cameron:1959}
A.~Cameron,
\newblock Astrophys. J. {\bf 130}, 916 (1959).

\bibitem{Zeldovich:1962}
Y.~Zeldovich,
\newblock Sov. Phys.-JETP {\bf 14}, 1143 (1962).

\bibitem{Lattimer:2012}
J.~Lattimer,
\newblock Annual Review of Nuclear and Particle Science {\bf 62}, 485 (2012).

\bibitem{Antoniadis:2013pzd}
J.~Antoniadis {\em et~al.},
\newblock Science {\bf 340}, 6131 (2013), arXiv:1304.6875.

\bibitem{Demorest:2010}
P.~Demorest, T.~Pennucci, S.~Ransom, M.~Roberts, and J.~Hessels,
\newblock Nature {\bf 467}, 1081 (2010), arXiv:1010.5788.

\bibitem{PhysRevLett.70.2220}
T.~Damour and G.~Esposito-Far\`ese,
\newblock Phys. Rev. Lett. {\bf 70}, 2220 (1993).

\bibitem{PhysRevD.54.1474}
T.~Damour and G.~Esposito-Far\`ese,
\newblock Phys. Rev. D {\bf 54}, 1474 (1996).

\bibitem{PhysRevLett.105.151102}
W.~C.~C. Lima, G.~E.~A. Matsas, and D.~A.~T. Vanzella,
\newblock Phys. Rev. Lett. {\bf 105}, 151102 (2010).

\bibitem{Landulfo:2014wra}
A.~G.~S. Landulfo, W.~C.~C. Lima, G.~E.~A. Matsas, and D.~A.~T. Vanzella,
\newblock Phys.Rev. {\bf D91}, 024011 (2015), arXiv:1410.2274.

\end{thebibliography}
\bibliographystyle{h-physrev}

\end{document}